\documentclass[english,11pt,a4paper]{article}
\usepackage{jcappub}

\usepackage[T1]{fontenc}
\usepackage[latin9]{inputenc}
\usepackage{amsmath}
\usepackage{graphicx}
\usepackage{amssymb}

\makeatletter
\@ifundefined{textcolor}{}
{%
 \definecolor{BLACK}{gray}{0}
 \definecolor{WHITE}{gray}{1}
 \definecolor{RED}{rgb}{1,0,0}
 \definecolor{GREEN}{rgb}{0,1,0}
 \definecolor{BLUE}{rgb}{0,0,1}
 \definecolor{CYAN}{cmyk}{1,0,0,0}
 \definecolor{MAGENTA}{cmyk}{0,1,0,0}
 \definecolor{YELLOW}{cmyk}{0,0,1,0}
 }

\usepackage{epsfig}
\usepackage{dcolumn}
\usepackage{bm}
\@ifundefined{definecolor}
 {\usepackage{color}}{}
\usepackage{verbatim}

\newcommand{\gsim}{\mbox{\raisebox{-.9ex}{~$\stackrel{\mbox{$>$}}{\sim}$~}}}

\def\thebiblio#1{
\begin{center}\bf \large References
\end{center}
\list
{[\arabic{enumi}]}{\settowidth\labelwidth{#1.}\leftmargin\labelwidth
 \advance\leftmargin\labelsep
 \usecounter{enumi}}
 \def\newblock{\hskip .11em plus .33em minus -.07em}
 \sloppy
 \sfcode`\.=1000\relax}

\makeatother

\usepackage{babel}
\begin{document}

\title{\boldmath
Eliminating the $\eta$-problem in SUGRA hybrid inflation with vector
backreaction}


\author[a]{Konstantinos Dimopoulos,}
\emailAdd{k.dimopoulos1@lancaster.ac.uk}

\author[b]{George Lazarides}
\emailAdd{lazaride@eng.auth.gr}

\author[a,b]{and\\
 Jacques M. Wagstaff}
\emailAdd{j.wagstaff@lancaster.ac.uk}

\affiliation[a]{Consortium for Fundamental Physics, %
Physics Department, Lancaster University, Lancaster LA1 4YB, U.K.}
\affiliation[b]{Physics Division, School of Technology, Aristotle University of Thessaloniki,\\
Thessaloniki 54124, Greece}

\arxivnumber{1111.1929}

\date{\today}
\abstract{
It is shown that, when the inflaton field modulates the gauge kinetic function
of the gauge fields in supergravity realisations of inflation, the dynamic
backreaction leads to a new inflationary attractor solution, in which the
inflaton's variation suffers additional impedance. As a result, slow-roll
inflation can naturally occur along directions of the scalar potential which
would be too steep and curved to support it otherwise. This provides a generic
solution to the infamous eta-problem of inflation in supergravity. Moreover,
it is shown that, in the new inflationary attractor, the spectral index of the
generated curvature perturbations is kept mildly red despite eta of order
unity. The above findings are applied to a model of hybrid inflation in
supergravity with a generic K\"{a}hler potential. The spectral index of the
generated curvature perturbations is found to be 0.97 - 0.98, in excellent
agreement with observations. The gauge field can play the role of the vector
curvaton after inflation but observable statistical anisotropy requires
substantial tuning of the gauge coupling.}
\maketitle
\flushbottom


\section{Introduction}

Cosmic Inflation is considered by most theoretical cosmologists as an integral
part of the so-called concordance model of cosmology, which complements the
standard hot big bang. 
Inflation not only overcomes the flatness and horizon problems but, crucially,
it provides an elegant mechanism for the generation of
the primordial curvature perturbation, which is responsible for the origin of
structure in the Universe \cite{book}.
Because of this, in the last 25 years there has been
a massive effort to design compelling models of inflation based on realistic
particle theory. This endeavour was accompanied with the influx of
observational data on cosmological perturbations with ever increasing
precision, which have the power to discriminate between inflation models.
However, despite the fact that observations are now putting
substantial pressure on inflation model-building, we are still far from
ascertaining which inflation model is the most likely, since no candidate
has been found with a clear advantage over others.

In the early years of inflation model-building emphasis was put on economy.
This gave birth to the inflationary paradigm which considered that inflation
and the cosmological perturbations are determined by a single degree of
freedom, a scalar field called the inflaton. This was not only because
inflation had to outperform the rival paradigm of cosmic
strings\footnote{Cosmic strings had an important advantage over
inflation in those days, because they could explain the magnitude of the
cosmological perturbations $\sim 10^{-5}$.} but also because it was the
simplest choice, which allowed a handle over calculations and provided
inflation with predictive power. Conceptual problems regarding initial
conditions were overcome by the no-hair theorem (which put the issue beyond our
reach) and by the idea of eternal inflation \cite{eternal},
which is an initial phase allowed
in many inflationary models. However, insisting on a single degree of freedom
resulted unavoidably to massive fine-tuning, which was still far less severe
than that of the horizon and flatness problems, but nevertheless, it was found
to plague the vast majority of inflation models.

One of the sources of fine-tuning was the apparent requirement of
super-Planckian values of the inflaton field in large-field inflation
models.\footnote{The small field models suffered from another fine-tuning
problem since the hilltop in the scalar potential had to be extremely flat.}
Super-Planckian field displacements implied that non-renormalisable terms in
the scalar potential could not be ignored. In fact, there was real danger of
them blowing up, rendering the theory out of control.

A seminal step to overcome the fine tunings of single-field inflation was made
by Linde in ref.~\cite{hybrid}, where he introduced a second scalar field (the
so-called waterfall field) interacting with the inflaton and causing the
termination of the inflationary phase. The model was named Hybrid Inflation and
is one of the most promising types of inflation models that remain into effect
today. In Hybrid Inflation the inflaton is relieved from one of its
responsibilities (to end inflation) and this is why the tuning requirements of
the model are reduced, at the expense of adding another degree of freedom. This
enabled the inflaton field to stay
sub-Planckian during inflation. Moreover, since inflation had to take place
at energies near the breaking of grand unification (to satisfy the
observational requirement on the magnitude of the curvature perturbation), the
waterfall field could be identified with the Higgs field of a Grand Unified
Theory (GUT).

However, the original Hybrid Inflation model suffered also from the problem of
excessive radiative corrections, which lifted the flatness of the inflaton
direction. To
harness this effect an additional symmetry was needed. Global supersymmetry
(SUSY) was proposed in ref.~\cite{susyhybrid}. SUSY offers a multitude of flat
directions whose flatness is guaranteed by the non-renormalisation theorem.
This is why SUSY has been extensively used in inflation model-building. In
SUSY Hybrid Inflation the radiative corrections are actually beneficial because
they provide a gentle logarithmic slope for the inflaton to slow-roll.

However, local supersymmetry, called supergravity (SUGRA), introduced another
set of fine-tunings because it turns out that K\"{a}hler corrections to the
scalar potential generically give rise to masses of order the Hubble scale $H$
to all scalar fields in the theory \cite{randall}. This is the basis for the
infamous $\eta$-problem of SUGRA inflation, which has to do with the fact that
the slow-roll parameter $\eta$ was pushed to order unity by SUGRA corrections,
where
$$
\eta\equiv m_P^2\frac{V''}{V}\simeq\frac13\left(\frac{m}{H}\right)^2=
{\cal O}(1)\,,
$$
with \mbox{$m\sim H$} being the mass of the scalar field, the prime denotes
derivative of the scalar potential $V$ with respect to the scalar field and
we used that the Friedmann equation during inflation is
\mbox{$3m_P^2H^2\simeq V$}, since the Universe is dominated by the
potential density of the inflaton. Note here that the above argument is true
for any light scalar field during inflation; not only the inflaton. This is a
problem because, if a scalar field contributes to the curvature perturbation
then the spectral index $n_s$ receives from $\eta$ a contribution of the form
$$
\delta(n_s-1)=2\eta\,.
$$
Thus, to obtain the approximate scale invariance revealed by the
CMB observations, one requires \mbox{$\eta\ll 1$}. A second problem has to do
with the requirement that the inflaton slow-rolls, since we need about 60
e-folds of inflation to solve the horizon and flatness problems. However,
SUGRA corrections lift the flatness of the inflaton direction and destabilise
slow-roll. Finally, a third problem is that, if the scalar field mass is
\mbox{$m\gsim\frac32 H$},
particle production is exponentially suppressed, so there is danger of not
being able to generate the density perturbations using scalar fields.

For Hybrid Inflation, the above does not affect the waterfall field whose mass
is generally much larger (except momentarily near the end of inflation), but
it could potentially affect the inflaton direction. Fortunately, when a minimal
K\"{a}hler potential is assumed, a cancelation protects the
model from the excessive supergravity corrections, and allows it to escape
the $\eta$-problem \cite{sugrahybrid}.
Yet, the price to pay was that a minimal K\"{a}hler
potential {\em has} to be assumed because any generic K\"{a}hler potential
(or any higher order corrections to the minimal K\"{a}hler potential)
would produce a massive $\eta$-problem and render the model nonviable.

In the meantime the observation of the CMB acoustic peaks lead to the collapse
of the cosmic string paradigm and put inflation at the centre-stage as the
most compelling paradigm for the origin of structure. Gradually, model-builders
began to move away from single-field inflation models of the old days and
started designing more complicated models with better connection to particle
theory. It was soon realised that adding new degrees of freedom reduces the
fine-tuning requirements of inflation models. Thus, there was a wave of new
models involving many scalar fields such as double inflation \cite{double},
extended inflation \cite{extended} and more recently
assisted inflation \cite{assisted} (or N-flation \cite{Nflation}), where a
cascade of hundreds of identical scalar fields is assumed.
Another approach mirrored the idea of Hybrid Inflation in that it removed the
responsibility from the inflaton to accomplish
some of its tasks, assigning it to another field. In the curvaton paradigm,
for example, the inflaton is no more required to generate the curvature
perturbation through its quantum fluctuations. Instead this task is
given to some other so-called curvaton field \cite{curv}, which is unrelated to
the physics of inflation. Many curvaton candidates were found in simple
extensions of the standard model of particle physics, while it was demonstrated
that inflation model-building was liberated by the introduction of a curvaton
field \cite{liber}. However, being a scalar field, the curvaton also suffers
from the $\eta$-problem and the same is true for all other proposals which
include many scalar fields.\footnote{The exception is pseudo-Nambu Goldstone
bosons (PNGBs), whose flatness is preserved by the remnant of the
shift-symmetry. Natural inflation \cite{natural} exploits this advantage, but
unfortunately, to produce the observed curvature perturbation, the order
parameter of the PNGB has to be super-Planckian, in contrast to
quantum gravity / string theory considerations. Recently, a promising way out
was found in ref.~\cite{lorenzo} by gauging the PNGB inflaton. In this way,
vector field backreaction from the axionic coupling facilitates steep
inflation, which allows the order parameter (decay constant) of the inflaton to
be sub-Planckian. This proposal has parallels to our work
(in the sense that vector backreaction allows steep inflation),
even though PNGBs do not suffer from the $\eta$-problem {\em per se}.
For a PNGB curvaton see ref.~\cite{pngb}.}

Recently a surprising solution to the $\eta$-problem of SUGRA inflation
was discovered. In ref.~\cite{Wagstaff} it was shown that an interaction
between the inflaton and a vector boson field affects profoundly the inflaton's
evolution. Indeed, a new inflationary attractor solution was found (called the
vector scaling solution) where the vector field backreaction ${\cal B}_A$
impedes the inflation's variation in that
it reduces the effective potential slope as experienced by the inflaton field:
\mbox{$|V'_{\rm eff}|<|V'|$}, where \mbox{$V'_{\rm eff}\equiv V'+{\cal B}_A$}
and the prime denotes derivative with respect to the inflaton field. This
allows steep inflation and can overcome the $\eta$-problem by enabling
slow-roll to take place even if the scalar potential is substantially curved.
Furthermore, the vector backreaction affects the inflaton equation of motion
such that it allows the inflaton to undergo particle production even with
\mbox{$\eta={\cal O}(1)$}.

In this paper, we demonstrate that the mechanism of vector backreaction not
only allows long-lasting slow-roll inflation with a steep and curved potential
but also protects the spectral index of the inflaton's perturbations against
excessive contributions from a large $\eta$ parameter. Thereby, vector
backreaction eliminates all aspects of the $\eta$-problem, the excessive tilt
of the spectrum and the destabilisation of slow-roll for the inflaton. We apply
these findings to the standard SUGRA Hybrid Inflation model with a generalised
K\"{a}hler potential, which goes beyond the minimal form. We find that
even with a non-minimal K\"{a}hler potential, Hybrid Inflation can be
long-lasting and produce a weakly red spectrum of curvature perturbations, in
agreement with observations.

Since the pioneering work in ref.~\cite{vecurv}, which introduced the vector
curvaton paradigm, cosmic vector fields are increasingly being considered to
affect the dynamics of inflation and the generation of the curvature
perturbation $\zeta$. Indeed, it was found that, the contribution of vector
fields can give rise to statistical anisotropy in $\zeta$
\cite{soda,stanis}. This is a new observable, which cannot be produced by
scalar fields alone as the latter cannot generate a preferred direction on the
microwave sky. In fact, a preferred direction might be hinted by the unlikely
correlation of the low multiples of the CMB \cite{AoE}. The observations still
allow as much as 30\% statistical anisotropy in the spectrum of the curvature
perturbation \cite{GE}. As for the bispectrum, this can be predominantly
anisotropic even if statistical anisotropy in the spectrum is small
\cite{fnlanis,bartolo}. The preferred direction of the statistical anisotropy
in the spectrum and bispectrum of the curvature perturbation is the same, which
is a smoking gun for the contribution of a vector field to $\zeta$
\cite{fnlanis}.

In ref.~\cite{Wagstaff} it was demonstrated that, when the inflaton affects the
kinetic function of an Abelian vector boson field, the new attractor solution
may render the Universe expansion mildly anisotropic (see also
ref.~\cite{mota}) and is such that it generates
scale-invariant spectra for the vector field components.
In refs.~\cite{anisinf, anisinf+}
it was shown that, anisotropic inflation gives rise to statistical anisotropy
in the perturbations of the inflaton field which, in turn, pass it over to
$\zeta$. Alternatively, one can employ the vector field perturbations
themselves to generate a statistically anisotropic $\zeta$ \cite{stanis}, e.g.
if the vector field acts as a vector curvaton as shown in
refs.~\cite{sugravec, varkin} (see also refs.~\cite{nonmin, stanis}).
Both contributions are complementary.

In this paper we apply the above in the generation of
statistical anisotropy in the spectrum and bispectrum of the curvature
perturbation in SUGRA Hybrid Inflation, by coupling the vector field to the
waterfall field so that it can
play the role of vector curvaton. Such a coupling was investigated in
ref.~\cite{soda} to generate statistical anisotropy in $\zeta$, through the
end-of-inflation mechanism, since the vector field perturbs the
moment when inflation is terminated by the breaking of grand unification. Our
investigation is complementary to that work as we study the additional
statistical anisotropy that can be generated through the vector curvaton
mechanism.

In our paper we consider natural units, where \mbox{$c=\hbar=k_B=1$} and
Newton's gravitational constant is \mbox{$8\pi G=m_P^{-2}$}, with
\mbox{$m_P=2.44\times10^{18}\text{GeV}$} being
the reduced Planck mass.

\section{Vector Scaling Slow-Roll Inflation}

Consider the \mbox{$U(1)$} gauge symmetry
\mbox{$\Phi\rightarrow\Phi e^{i\Lambda(x)}$}, where the group parameter
$\Lambda(x)$ is a function of spacetime coordinates $x$, acting on the
complex scalar field $\Phi$ with unit charge. To achieve invariance of the
kinetic term \mbox{$|D_{\mu}\Phi|^2$}, we need to include the Abelian gauge field $B_{\mu}$ with transformation \mbox{$h_0B_{\mu}\rightarrow h_0B_{\mu}+\partial_{\mu}\Lambda(x)$} via the covariant derivative \mbox{$D_{\mu}\equiv\partial_{\mu}-ih_0B_{\mu}$}, where $h_0$ is the gauge coupling constant. Switching to polar coordinates \mbox{$\Phi=\phi e^{i\theta}/\sqrt2$} and defining the (gauge invariant) combination \mbox{$h_0A_{\mu}\equiv h_0B_{\mu}-\partial_{\mu}\theta$}, we can obtain the following Lagrangian density for an Abelian Higgs model
\begin{equation}
 {\cal \mathcal{L}}=-\frac{1}{4}f(\sigma)F_{\mu\nu}F^{\mu\nu}
 +\frac12 \left(\partial_{\mu}\phi\right)\left(\partial^{\mu}\phi\right)
 +\frac{1}{2}h_{0}^{2}\phi^{2}A_{\mu}A^{\mu}
 -V_1(\phi),
\label{model}
\end{equation}
where the field strength tensor is $F_{\mu\nu}=\partial_{\mu}B_{\nu}-\partial_{\nu}B_{\mu}=\partial_{\mu}A_{\nu}-\partial_{\nu}A_{\mu}$ and $V_1(\phi)$ is the scalar potential for the Higgs field $\phi$. In this model the gauge kinetic function $f$ is modulated by another scalar field \mbox{$\sigma$} which will drive a period of inflation (this is natural in supergravity theories where $f$ is a holomorphic function of the scalar fields of the model). Defining the gauge kinetic function as follows
\begin{equation}
 f(\sigma)\equiv\left(\frac{h_{0}}{h(\sigma)}\right)^{2},
\end{equation}
where we assume that \mbox{$h(\sigma_0)=h_0$} so that \mbox{$f(\sigma_0)=1$} and the vector field becomes canonically normalised when $\sigma$ assumes its vacuum expectation value (VEV) $\sigma_0$.

The mass of the gauge field, \mbox{$m\equiv h_0\phi$}, is given through the Higgs mechanism by the non-zero VEV $\phi_0$ of the Higgs field \mbox{$\phi$} which spontaneously breaks the gauge symmetry. The spatial components of the physical vector field are given by \mbox{$W_i=\sqrt{f}A_i/a$}, with mass \mbox{$M_A\equiv m/\sqrt{f}=h(\sigma)\phi$}. Hence, once the gauge symmetry is spontaneously broken
the physical vector field obtains a mass \mbox{$m_A\equiv M_A(\phi_0,\sigma_0)=h_0\phi_0$}, where we assumed that \mbox{$\sigma\rightarrow\sigma_0$} after the phase transition, i.e. the phase transition terminates inflation, as is usually the case in Hybrid Inflation.

It was shown in ref.~\cite{vecurv} that, as inflation homogenises the vector field \mbox{$\partial_iA_{\mu}=0$}, the temporal component vanishes \mbox{$A_0=0$} (if \mbox{$m=0$}, we can set \mbox{$A_0=0$} by a gauge choice). Thus, without loss of generality, the spatial components of the vector field can be lined up in the z axis \mbox{$A_{\mu}=(0,0,0,A_z(t))$}. We can then assume a Bianchi-I background with residual isotropy in the plane perpendicular to the vector expectation value \mbox{$ds^2=dt^2-a_1^2(t)(dx^2+dy^2)-a_2^2(t)dz^2$}, where $a_{1,2}$ are the scale factors related to the different spatial directions. We can then define the average scale factor \mbox{$a\equiv(a_1^2a_2)^{1/3}$} and the average Hubble rate as \mbox{$H\equiv\dot{a}/a$}, where a dot denotes a derivative with respect to cosmic time~$t$. The anisotropic stress $\Sigma$ induced by the vector field is then given by
\begin{equation}\label{stress-definition}
    \Sigma\equiv\frac{1}{3H}\frac{d}{dt}\ln\left(\frac{a_1}{a_2}\right).
\end{equation}

The coupling of the vector field to the scalar field, through the kinetic function \mbox{$f(\sigma)$}, induces a source term \mbox{$\mathcal{B}_A\equiv - a_2^{-2}f'(\sigma)\dot{A}^2_z/2$} in the scalar field equation
\begin{equation}
 \ddot{\sigma}+3H\dot{\sigma}+V'(\sigma)+\mathcal{B}_A(\sigma,\dot{A}_z)=0,
\end{equation}
where a prime denotes a derivative with respect to the field \mbox{$\sigma$} and $V(\sigma)$ is the scalar potential associated with the field $\sigma$.

As shown in ref.~\cite{Wagstaff}, for half of the model parameter space, and for sufficiently flat potentials, the system may evolve to the Standard Slow-Roll (SSR) inflationary attractor with $\sigma$ as the inflaton. On the SSR attractor the vector backreaction \mbox{$\mathcal{B}_A$} and the anisotropic stress $\Sigma$ vanish. The vector field energy density $\rho_A$ also vanishes on the SSR attractor and hence cannot influence the expansion of the Universe. The SSR attractor provides the typical example to the cosmic no-hair theorem \cite{wald}.

For the other half of parameter space, the source term \mbox{$\mathcal{B}_A$} can backreact on the dynamics of the inflaton field and lead to novel attractor solutions. In this case the vector backreaction grows and dynamical analysis shows that solutions tend to the Vector Scaling Slow-Roll (VSSR) inflationary attractor \cite{Wagstaff}.\footnote{A similar stability analysis for the massless version of our model in ref.~\cite{Wagstaff} assuming an exponential functional dependence for $f(\sigma)$ and $V(\sigma)$ was performed in refs.~\cite{sodanew} and \cite{mota}.} On the VSSR attractor the vector field has a non-negligible effect on the expansion of the Universe through its non-vanishing energy density and hence non-vanishing anisotropic stress. The VSSR attractor, being an anisotropic inflationary solution, nicely provides a counter example to the cosmic no-hair theorem \cite{wald}.

On the VSSR attractor, as was shown in ref.~\cite{Wagstaff}, the gauge kinetic function scales as \mbox{$f(\sigma)=f_{\text{att}}\propto a^{-4}$} (we notice that because \mbox{$f\propto 1/h^2$} and \mbox{$f\rightarrow1$} at the end of inflation, the gauge field remains weakly coupled throughout inflation). This scaling of the kinetic function leads to scale-invariant transverse spectra\footnote{%
Where `L' and `R' denote the left and right transverse polarizations respectively and `$\parallel$' denotes the longitudinal polarization. As this is a parity conserving theory the left and right components are identical.} \mbox{$\mathcal{P}_{L,R}$} of vector field perturbations \cite{varkin}. If the vector field is massless the longitudinal component and its perturbations are decoupled from the theory. In this case,
\mbox{$f\propto a^{-4}$} is all we need to consider, if we are interested in a significant effect of the vector field on cosmological scales (such as a contribution to the curvature perturbation of the Universe $\zeta$).\footnote{For a review of the vector curvaton paradigm and statistical anisotropy see ref.~\cite{vecrev}.} Because we are left with the transverse components only, particle production of the vector field is highly anisotropic. Therefore, the vector field can only contribute subdominantly to $\zeta$. However for a massive vector field, we have to consider also the longitudinal component, which turns out to generate the dominant contribution. To obtain a scale-invariant longitudinal spectrum of perturbations \mbox{$\mathcal{P}_{\parallel}$}, we also require that \mbox{$m\propto a$} \cite{varkin}. In this case the phenomenology is richer as the vector field spectra depend on the mass of the physical vector field at the end of inflation. If the field remains light until the end of inflation we have highly anisotropic vector field spectra and the field can only contribute subdominantly to $\zeta$, but it can still generate substantial statistical anisotropy and anisotropic non-Gaussianity in the primordial curvature perturbation \cite{varkin}. If the vector field becomes heavy by the end of inflation then its spectra become isotropic and the field can then alone generate $\zeta$ \cite{varkin}.

In ref.~\cite{Wagstaff}, the VSSR attractor solution, including the energy densities and anisotropic stress, was obtained in terms of the dimensionless model parameters defined by
\begin{equation}\label{eq:mod-parameters}
 \Gamma_{f}(\sigma)\equiv\sqrt{\frac{3}{2}}m_{P}\left(\frac{f'}{f}\right)
 \qquad\text{and}\qquad
 \lambda_{0}(\sigma)\equiv\sqrt{\frac{3}{2}}m_{P}\left(\frac{V'}{V}\right)\Biggr|_{\phi=0},
\end{equation}
where we are considering a massless gauge field, i.e. unbroken gauge symmetry where \mbox{$\phi=0$}. The backreaction criteria for the VSSR attractor, given in ref.~\cite{Wagstaff}, are simply
\begin{equation}
 \textrm{Conditions\quad}
  \begin{cases}
    {\rm I} & \quad|\Gamma_{f}|\gg1,\\
    {\rm II} & \quad|\Gamma_{f}|\gg|\lambda_{0}|,\\
    {\rm III} & \quad\lambda_{0}\Gamma_{f}>6.
  \end{cases}
\label{eq:conditions_1-2-3}
\end{equation}

The VSSR attractor is a stationary point in phase-space if the dimensionless model parameters are exactly constant i.e. for exponential forms of the kinetic function and scalar potential. However, for more general functional forms of the kinetic function and potential, the dimensionless model parameters are time-dependent. Therefore the VSSR attractor is really a moving point in phase-space. As shown in ref.~\cite{Wagstaff}, additional conditions arise to guarantee that solutions are able to reach the moving attractor and be dragged along with it. These conditions are given by
\begin{equation}\label{eq:CondA-D}
A,B,C,D
<1,
\end{equation}
where the functions \mbox{$A,B,C$} and $D$ are defined by
\begin{eqnarray}
 A & \equiv & 4\sqrt{\frac{2}{3}}m_{P}\left|\frac{\Gamma_{f}'}{\Gamma_{f}^{2}}
 +\frac{1}{3}\frac{\lambda_{0}'}{\Gamma_{f}^{2}}\right|,\label{eq:CondA-1}\\
 B & \equiv & 2\sqrt{\frac{2}{3}}m_{P}\left|\frac{\lambda_{0}'}{\Gamma_{f}^{2}}\right|,\\
 C & \equiv & 4\sqrt{\frac{2}{3}}m_{P}\left|\frac{\Gamma_{f}'}{\Gamma_{f}^{2}}
 +\frac{\lambda_{0}'}{\Gamma_{f}^{2}}-\frac{\lambda_{0}'\Gamma_{f}
 +\lambda_{0}\Gamma_{f}'}{\Gamma_{f}\left(\lambda_{0}\Gamma_{f}-6\right)}\right|,\label{eq:CondC}\\
 D & \equiv & 4\sqrt{\frac{2}{3}}m_{P}\left|-\frac{2}{3}\frac{\lambda_{0}'}{\Gamma_{f}^{2}}
 +\left(\frac{\Gamma_{f}'}{\Gamma_{f}^{2}}\right)\left(\frac{2}{3}\frac{\lambda_{0}}{\Gamma_{f}}
 -\frac{8}{\Gamma_{f}^{2}}\right)\right|.\label{eq:CondD-1}
\end{eqnarray}

\subsection{General properties of the VSSR attractor}

In this section we summarise briefly the results obtained in ref.~\cite{Wagstaff} for the VSSR inflationary attractor.

The vector backreaction ${\cal B}_A\equiv - a_2^{-2}f'(\sigma)\dot{A}^2_z/2$ is not dynamic with respect to the scalar field since it is not a function of $\dot{\sigma}$. Thus the backreaction may be interpreted as to only modify the effective slope of the potential \mbox{$V'_{\text{eff}}\equiv V'+\mathcal{B}_A$}. As we are considering gauge symmetry the gauge kinetic function \mbox{$f\propto1/h^2$}, and we assume canonical normalisation \mbox{$f\rightarrow1$} at the end of inflation, we require that the kinetic function is always decreasing in time, \mbox{$\dot{f}(t)<0$}, so that the gauge field remains weakly coupled. Because of this we notice that the backreaction ${\cal B}_A$ always has an opposite sign to the potential slope \mbox{$V'(\sigma)$}. This is because the scalar field rolls down its potential, so if \mbox{$V'(\sigma)>0$}, then \mbox{$\dot{\sigma}<0$} and therefore \mbox{$f'(\sigma)>0$}; however if \mbox{$V'(\sigma)<0$}, then \mbox{$\dot{\sigma}>0$} and therefore \mbox{$f'(\sigma)<0$}. This ensures that ${\cal B}_A$ always has an opposite sign to the potential slope \mbox{$V'(\sigma)$}. We conclude that the backreaction (if non-negligible) will always reduce the effective potential slope experienced by the inflaton and slow down the scalar field as it rolls down its potential. And for the same reasons as described above, the dimensionless model parameters $\lambda_0$ and $\Gamma_f$ always have the same sign and thus the combination \mbox{$\lambda_0\Gamma_f$} is always positive.

Once the VSSR attractor is obtained, which is a fixed point in phase-space (it may be a slowly moving point depending on the dimensionless model parameters), the backreaction becomes proportional to the potential slope \mbox{$V'(\sigma)$}. Therefore the effective scalar potential slope $V'_{\text{eff}}$ as seen by the scalar field is given by \cite{Wagstaff}
\begin{equation}\label{eq:V_eff}
 V'_{\textrm{eff}}\equiv V'+\mathcal{B}_A
\simeq\frac{6}{\lambda_{0}\Gamma_{f}}V'.
\end{equation}
Considering condition III in eq.~(\ref{eq:conditions_1-2-3}), we observe that the effective potential slope seen by the inflaton is reduced. This effect can be used to obtain slow-roll inflation in models with potentials that would normally be too steep to support a slow-roll regime. Indeed this attractor can be used to solve the infamous $\eta$-problem of supergravity (see section \ref{SUSY_Hybrid}). On the attractor the slow-roll parameters \mbox{$\epsilon_{\textrm{H}}\equiv-\dot{H}/H^2$} and \mbox{$\eta_{\textrm{H}}\equiv-\ddot{H}/2H\dot{H}$} are given by
\begin{equation}\label{eq:slow-roll_gen}
 \epsilon_{\textrm{H}}\simeq\frac{2\lambda_{0}}{\Gamma_{f}}\ll1
 \quad\text{and}\quad
 \eta_{\textrm{H}}\simeq\frac{2\lambda_{0}}{\Gamma_{f}}
 +\frac{\sqrt{6}m_{P}}{\Gamma_{f}}\left(\frac{\lambda_{0}'}{\lambda_{0}}
 -\frac{\Gamma_{f}'}{\Gamma_{f}}\right).
\end{equation}
Therefore a stage of vector scaling slow-roll inflation with \mbox{$\epsilon_{\textrm{H}},\eta_{\textrm{H}}\ll1$} is possible. Hence we may consider the slow-roll equations
\begin{equation}\label{eq: slow-roll eqs}
 3m_{P}^{2}H^{2}\simeq V(\sigma)\qquad\text{and}\qquad3H\dot{\sigma}\simeq-V'_{\text{eff}}(\sigma),
\end{equation}
where \mbox{$H$} is now the inflationary Hubble scale on the VSSR attractor.

The number of e-foldings of expansion $N_{\text{att}}$ during the period of VSSR inflation, where \mbox{$f(\sigma)=f_{\text{att}}(\sigma)\propto a^{-4}$}, is given by
\begin{equation}\label{eq:efoldings}
 N_{\text{att}}=\frac{1}{4}\ln\frac{f(\sigma_{i})}{f(\sigma_{\text{end}})},
\end{equation}
where we denote the field values at the start and end of the attractor as \mbox{$\sigma_{i}$} and \mbox{$\sigma_{\text{end}}$} respectively.

The vector-to-scalar field energy density ratio $\mathcal{R}$ does not vanish on the attractor, as it would normally do in the standard slow-roll case, but acquires a small non-zero value given by \cite{Wagstaff}
\begin{equation}\label{eq:RandSigma}
 \mathcal{R}\equiv\frac{\rho_{A}}{\rho_{\sigma}}\simeq\frac{\lambda_{0}\Gamma_{f}-6}{\Gamma_{f}^{2}}.
\end{equation}
This non-vanishing vector field energy density induces a small anisotropic stress, given by \mbox{$\Sigma\simeq2\mathcal{R}/3$}. The VSSR attractor therefore leads to mildly anisotropic inflation which can also source statistical anisotropy in the primordial curvature perturbation $\zeta$ \cite{anisinf,anisinf+}.

\section{The primordial curvature perturbation from VSSR inflation}
We will now calculate, for the first time, inflationary observables that arise from inflaton perturbations generated during the VSSR attractor. At horizon exit of the mode $k$ during inflation (in particular, the epoch of horizon exit for the pivot scale \mbox{$k=0.002\text{Mpc}^{-1}$} \cite{book}, which we denote by the subscript $*$), the curvature perturbation generated by inflaton perturbations is given by
\begin{equation}\label{eq:dp/p}
 \frac{2}{5}\zeta_{\sigma}=\frac{\delta\rho_{\sigma}}{\rho_{\sigma}}\Biggr|_{*}
 =\frac{1}{5\sqrt{3}\pi}\frac{V^{3/2}}{m_{P}^{3}\left|V'_{\textrm{eff}}\right|}\Biggr|_{*}
 \simeq\frac{1}{30\sqrt{2}\pi}\frac{V^{1/2}|\Gamma_{f}|}{m_{P}^{2}}\Biggr|_{*},
\end{equation}
where we used the slow-roll equations in eq.~(\ref{eq: slow-roll eqs}) together with eq.~(\ref{eq:V_eff}). Considering that the observed curvature perturbation $\zeta$ is generated by the slowly-rolling inflaton we have \mbox{$\zeta_{\sigma}\simeq\zeta$}, where the COBE normalisation is \mbox{$\zeta\simeq4.8\times10^{-5}$} \cite{wmap}. We notice that, on the VSSR attractor, the curvature perturbation depends on the inflationary scale and the ``slope'' (modulation rate $f'$) of the kinetic function, through \mbox{$\Gamma_{f}$}, but not on the slope of the potential \mbox{$V'$}.

\subsection{The spectral index and its running}

Considering that the entire curvature perturbation is generated by the inflaton, whose spectrum (assuming that it is a light field) is given by \mbox{$\mathcal{P}_{\sigma}=\left(\frac{H_{*}}{2\pi}\right)^{2}$}, we find \cite{book}
\begin{equation}
 \mathcal{P}_{\zeta}\simeq
 \frac{1}{4\pi^{2}}\left(\frac{H^{2}}{\dot{\sigma}}\right)^{2}\Biggr|_{*}.
\end{equation}
During the VSSR attractor we may use the slow-roll equations in eq.~(\ref{eq: slow-roll eqs}) evaluated at the time when cosmological scales exit the horizon, hence
\begin{equation}
 \mathcal{P}_{\zeta}\simeq\frac{1}{24\pi^{2}m_{P}^{4}}\frac{V(\sigma)}{\epsilon(\sigma)}
 \left(\frac{\lambda_{0}\Gamma_{f}}{6}\right)^{2}\Biggr|_{*},
\end{equation}
where the slow-roll parameters are defined in the usual way \mbox{$\epsilon(\sigma)\equiv (m_P^2/2)\left(V'/V\right)^2$} and \mbox{$\eta(\sigma)\equiv m_P^2\left(V''/V\right)$}. The spectral index is given by
\begin{equation}
 n_{s}-1\equiv\frac{d\ln\mathcal{P}_{\zeta}}{d\ln k}
 =\frac{\dot{\sigma}}{H}\frac{1}{\mathcal{P}_{\zeta}}\frac{d\mathcal{P}_{\zeta}}{d\sigma}\Biggr|_{*},
\end{equation}
where we used \mbox{$\frac{d}{d\ln k}=\frac{\dot{\sigma}}{H}\frac{d}{d\sigma}$} in the last equality. On the VSSR attractor the spectral index becomes
\begin{eqnarray}
 n_{s}-1 & \simeq & \left(\frac{6}{\lambda_{0}\Gamma_{f}}\right)\left[2\eta-6\epsilon-2m_{P}\sqrt{\frac{2}{3}}
 \frac{\left(\lambda_{0}\Gamma_{f}\right)'}{\Gamma_{f}}\right],\\
 & = & -2\left(\frac{6}{\lambda_{0}\Gamma_{f}}\right)
 \left[\epsilon+m_{P}\sqrt{\frac{2}{3}}\frac{\lambda_{0}\Gamma_{f}'}{\Gamma_{f}}\right].
 \label{eq: spectral index}
\end{eqnarray}
This result is completely general for the attractor solution, we have not yet specified any model. We recover the standard result \mbox{$n_{s}-1=2\eta-6\epsilon$} for \mbox{$V'_{\text{eff}}=V'$}, i.e. \mbox{$\lambda_{0}\Gamma_{f}=6$}, hence \mbox{$\left(\lambda_{0}\Gamma_{f}\right)'=0$}.

We notice that the spectral index becomes independent of the potential curvature encoded in $\eta$. It is therefore easy to obtain a red spectrum independently of the curvature. In effect the spectral index now depends on the ``curvature'' of the kinetic function ($f''$) through $\Gamma_f'$.

The current CMB observational bounds from WMAP7 on the scalar spectral index are \mbox{$0.953\leq n_s\leq0.981$} and \mbox{$n_s - 1 = -0.033 \pm 0.014$} (at 1$\sigma$) \cite{wmap}, i.e. they favour a red spectrum, although exact scale-invariance is not ruled out.

The running of the spectral index, given by
\mbox{$n_{s}'\equiv\frac{dn_{s}}{d\ln k}$}, on the VSSR attractor becomes
\begin{multline}
 n_{s}'\simeq\left(\frac{6}{\lambda_{0}\Gamma_{f}}\right)^{2}\Bigg\{16\epsilon\eta-24\epsilon^{2}-2\xi\\
 -\sqrt{\frac{2}{3}}m_{P}\frac{\left(\lambda_{0}\Gamma_{f}\right)'}{\Gamma_{f}}\left[6\epsilon
 -2\eta+2\sqrt{\frac{2}{3}}m_{P}\frac{\left(\lambda_{0}\Gamma_{f}\right)'}{\Gamma_{f}}\right]
 +\frac{4}{3}m_{P}^{2}\frac{\lambda_{0}}{\Gamma_{f}}\left[\left(\lambda_{0}\Gamma_{f}\right)''
 -\frac{\Gamma_{f}'\left(\lambda_{0}\Gamma_{f}\right)'}{\Gamma_{f}}\right]\Bigg\},
 \label{eq:running1}
\end{multline}
where \mbox{$\xi\equiv m_P^4 \left(V'V'''/V^2\right)$}. The standard result \mbox{$n_{s}'=16\epsilon\eta-24\epsilon^{2}-2\xi$} is again recovered for \mbox{$V'_{\text{eff}}=V'$}, i.e. \mbox{$\lambda_{0}\Gamma_{f}=6$}, hence \mbox{$\left(\lambda_{0}\Gamma_{f}\right)'=0$}. The result in eq.~(\ref{eq:running1}) can be written as
\begin{equation}\label{eq:running}
 n_{s}'\simeq2\epsilon\left(\frac{6}{\lambda_{0}\Gamma_{f}}\right)^{2}\left\{ \eta-2\epsilon+2m_{P}^{2}\left[\frac{\Gamma_{f}''}{\Gamma_{f}}
 -2\left(\frac{\Gamma_{f}'}{\Gamma_{f}}\right)^{2}\right]\right\}.
\end{equation}

The current CMB observational bounds, from WMAP7, on the running of the spectral index with no gravitational waves are \mbox{$-0.084 < n_s' < 0.020$} (at \mbox{$95\%\text{cf}$}) \cite{wmap}.

\subsection{The tensor spectrum}

We should also consider the primordial tensor perturbation which may be generated during inflation. Its spectrum $\mathcal{P}_{h}$ and corresponding spectral index $n_{T}$ are given by \cite{book}
\begin{equation}
 \mathcal{P}_{h}(k)=\frac{8}{m_{P}^{2}}\left(\frac{H_{*}}{2\pi}\right)^{2}
 \qquad\text{and}\qquad
 n_{T}\equiv\frac{d\ln\mathcal{P}_{h}(k)}{d\ln k}=-2\epsilon_{\text{H}}.
\end{equation}
An important quantity is the tensor-to-scalar ratio defined as \mbox{$ r\equiv\mathcal{P}_{h}/\mathcal{P}_{\zeta}$}. On the VSSR attractor we find the following results
\begin{equation}\label{eq:Tensor-scalar}
 r\simeq16\epsilon\left(\frac{6}{\lambda_{0}\Gamma_{f}}\right)^{2}=\frac{192}{\Gamma_{f}^{2}}
 \qquad\text{and}\qquad n_{T}\simeq-\frac{4\lambda_{0}}{\Gamma_{f}}=-\frac{r}{8}\left(\frac{\lambda_{0}\Gamma_{f}}{6}\right).
\end{equation}
Hence the `consistency relation' of standard slow-roll inflation, \mbox{$n_T=-r/8$}, does not hold for VSSR inflation. Therefore, if $r$ and $n_T$ are observed and do not satisfy the `consistency relation', the evidence could point towards other inflationary attractors such as the VSSR attractor.

The current CMB observational bound from WMAP7 on gravitational waves with no running is \mbox{$r < 0.36$} (at \mbox{$95\%\text{cf}$}) \cite{wmap}.

\section{Statistical anisotropy from the vector curvaton}

Clearly, because of the non-vanishing vector field energy density during VSSR inflation \mbox{$\mathcal{R}\neq0$}, we have to consider the possibility that vector field perturbations contribute to the primordial curvature perturbation of the Universe \cite{vecurv,sugravec}. The vector field must be light during inflation for successful particle production. But the vector field cannot dominate the energy density as it would cause excessive anisotropic expansion in conflict with observations (notice how this condition is easily achieved through the VSSR attractor predicting \mbox{$\mathcal{R}\ll1$}). However, if the vector field becomes heavy after inflation it begins to oscillate rapidly and behaves like pressureless matter \cite{vecurv}. The vector field can then (nearly) dominate the energy density and imprint its spectra of perturbations through the vector curvaton mechanism~\cite{vecurv}.

The simplest possibility is that the vector field remains massless during
inflation. Indeed,
consider first that inflation occurs while \mbox{$\phi=0$}, therefore the vector field is massless. The spectra of perturbations generated during inflation for a massless vector field is given by \cite{varkin}
\begin{equation}\label{eq:spectra}
 \mathcal{P}_{L,R}=\mathcal{P}_{\sigma}
 =\left(\frac{H_{*}}{2\pi}\right)^{2}\qquad\text{and}\qquad\mathcal{P}_{\parallel}=0.
\end{equation}
These scale-invariant results for the transverse components of the vector field spectra are obtained only if \mbox{$f\propto a^2$} or \mbox{$f\propto a^{-4}$} \cite{varkin}. As seen, the latter condition arises automatically out of the VSSR attractor~\cite{Wagstaff}.

To act as a vector curvaton the vector field needs to obtain a non-zero mass.
This is achieved by Higgsing the vector field as shown in eq.~(\ref{model}).
The Higgs field assumes a non-zero expectation value at the phase transition
which terminates inflation. This in turn generates a mass for the vector field.
Once the vector field obtains a mass, by the non-zero VEV of the Higgs field $\phi_0$ which spontaneously breaks the gauge symmetry, it may undergo coherent oscillations. The vector field then has a particle interpretation and its decay rate $\Gamma_{A}$ is given by
\begin{equation}\label{eq:Gamma_A}
 \Gamma_{A}=\frac{h_{0}^{2}M_{A}}{8\pi}=\frac{h_{0}^{3}\phi_{0}}{8\pi}=H_{\text{dec}},
\end{equation}
where the subscript `dec' denotes the epoch of vector field decay. Let us define the following useful parameter
\begin{equation}\label{eq:varepsilon-def}
 \varepsilon\equiv\frac{\Gamma_{A}}{H_{\text{end}}}
 \simeq \frac{h_0^3\phi_0|\Gamma_{f}(\sigma_{*})|}{32\sqrt{6}\pi^2m_{P}\zeta},
\end{equation}
where eqs.~(\ref{eq:dp/p}) and (\ref{eq:Gamma_A}), together with the assumption that \mbox{$H_{\text{end}}\simeq H_*$} (where the subscript `end' marks the end of inflation), were used to obtain the second equality. For the gravitational effect of the vector curvaton not to be suppressed we require that \mbox{$\varepsilon\lesssim1$}.\footnote{%
If it is to affect the Universe expansion, the oscillations of the vector curvaton should last at least a Hubble time before its decay.}
From this requirement, an upper bound on the gauge coupling constant is obtained
\begin{equation}\label{eq:h0_bound1}
 h_{0}\lesssim\left(\frac{32\sqrt{6}\pi^{2}m_{P}\zeta}{\phi_0|\Gamma_{f}(\sigma_{*})|}
 \right)^{1/3}.
\end{equation}

From eq.~(\ref{eq:spectra}) we see that the vector field spectra of perturbations are completely anisotropic \mbox{$\mathcal{P}_{L,R}\neq\mathcal{P}_{\parallel}$}, and therefore cannot be the dominant contribution to $\zeta$. The contribution of vector field perturbations will be in generating statistical anisotropy. Statistical anisotropy in the spectrum can be parameterized as
\begin{equation}
    \mathcal{P}_{\zeta}(\mathbf{k})
    =\mathcal{P}^{\text{iso}}_{\zeta}(k)
    \left[ 1+g\left(\mathbf{\hat{d}\cdot\hat{k}}\right)^2+\cdots \right],
\end{equation}
where \mbox{$\mathbf{\hat k}\equiv \mathbf k/k$}, $\mathbf{\hat d}$ is the unit vector in the preferred direction chosen by the homogenised vector field, $g$ quantifies the statistical anisotropy in the spectrum and the ellipsis denotes higher order terms which are not important if \mbox{$g<1$}.

In ref.~\cite{GE} it was found that \mbox{$g=0.29\pm 0.03$} at the level of $9\sigma$. However, the preferred direction was too close to the ecliptic plane so the authors suspected some unknown systematic. Hence, this number can only be considered as an upper bound \mbox{$\left|g\right|\lesssim0.3$}. The observations of the Planck satellite will decrease this bound to \mbox{$\left|g\right|\lesssim0.02$} if it is not observed \cite{planck}.

For the vector curvaton case, in the so-called $\delta N$-formalism \cite{dN}, it was found in ref.~\cite{stanis} that
\begin{equation}\label{eq:g}
 g=\beta\frac{\mathcal{P}_{\parallel}-\mathcal{P}_{+}}{\mathcal{P}_{\sigma}
 +\beta\mathcal{P}_{+}}=\frac{-\beta}{1+\beta}\simeq-\beta,
\end{equation}
where \mbox{$\beta\equiv\left(N_{A}/N_{\sigma}\right)^{2}\ll 1$}, \mbox{$N_{\sigma}\equiv\frac{\partial N}{\partial \sigma}$} and \mbox{$N_A=|\mathbf{N_A}|$} where \mbox{$N_A^i\equiv\frac{\partial N}{\partial W_i}$}. Therefore we must now consider both the inflaton and the vector curvaton contributions to the primordial curvature perturbation. In the $\delta N$-formalism, to first order in perturbations, the total curvature perturbation is given by
\begin{equation}\label{eq:deltaN}
 \zeta=N_{\sigma}\delta\sigma+N_{A}\delta W.
\end{equation}
The total curvature perturbation can also be calculated as the sum of individual curvature perturbations multiplied by an appropriate weighting factor
\begin{equation}\label{eq:curvpertdecomp}
 \zeta=\left(1-\hat{\Omega}_{A}\right)\zeta_{\sigma}+\hat{\Omega}_{A}\zeta_{A},
\end{equation}
where
\begin{equation}
 \hat{\Omega}_{A}\equiv\frac{\rho_{A}+p_{A}}{\rho+p}
 \qquad\text{and}\qquad
 \zeta_{i}=-H\frac{\delta\rho_{i}}{\dot{\rho}_{i}},
\end{equation}
where the subscript `$i$' labels the different components of the Universe content and $\rho$ and $p$ is the total energy density and pressure respectively. We evaluate the vector field curvature perturbation when it decays, i.e. when it is oscillating and the average vector field pressure \mbox{$\bar{p}_{A}=0$} \cite{vecurv}. Therefore we obtain
\begin{equation}\label{eq:Omega_A}
 \hat{\Omega}_{A}=\frac{\mathcal{R}_{\text{dec}}}{\mathcal{R}_{\text{dec}}+n},
\end{equation}
where \mbox{$n\equiv1+p_{\sigma}/\rho_{\sigma}$}. The vector field may decay either when the scalar field is still oscillating i.e. \mbox{$\bar{p}_{\sigma}=0$} (where $n=1$) or during the following stage of radiation domination, i.e. \mbox{$p_{\sigma}=p_{\rm rad}=\rho_{\rm rad}/3$} (where $n=4/3$). From eqs.~(\ref{eq:deltaN}) and (\ref{eq:curvpertdecomp}), we obtain
\begin{equation}\label{eq:NaNs}
 \frac{N_{A}\delta W}{N_{\sigma}\delta\sigma}=\frac{\hat{\Omega}_{A}}{1-\hat{\Omega}_{A}}\frac{\zeta_{A}}{\zeta_{\sigma}}
 =\frac{\mathcal{R}_{\text{dec}}}{n}\frac{\zeta_{A}}{\zeta_{\sigma}}\,.
\end{equation}
Then, using that the typical perturbation for the fields is given by
\mbox{$\delta W_{L,R}=\delta\sigma=\left(H_*/2\pi\right)$} we find the statistical anisotropy
\begin{equation}\label{eq:rootg1}
 \sqrt{\left|g\right|}\simeq\frac{N_{A}}{N_{\sigma}}
 =\frac{\mathcal{R}_{\text{dec}}}{n}\frac{\zeta_{A}}{\zeta_{\sigma}}.
\end{equation}
Let us now calculate the individual curvature perturbations \mbox{$\zeta_{i}=-H\left(\delta\rho_i/\dot{\rho}_i\right)$}. For the inflaton this becomes \mbox{$\zeta_{\sigma}\simeq\zeta$}, as the vector field can only contribute subdominanty to generate statistical anisotropy. For an oscillating vector field we have \cite{vecurv}
\begin{equation}\label{eq:zeta_A}
 \zeta_{A}=\frac{\delta\rho_{A}}{3\rho_{A}}\Biggr|_{\text{dec}}
 \simeq\frac{2}{3}\frac{\delta A}{A}\Biggr|_{\text{dec}}
 \approx\frac{H_{*}}{3\pi W_{\text{osc}}},
\end{equation}
where $W_{\text{osc}}$ is the value of the homogeneous physical vector field at the onset of oscillations and `dec' denotes the time of decay of the vector field. Hence, from eq.~(\ref{eq:rootg1}) we obtain
\begin{equation}\label{eq:g-general1}
 \sqrt{\left|g\right|}\approx
 \frac{H_{*}\mathcal{R}_{\text{dec}}}{3n\pi\zeta W_{\text{osc}}}.
\end{equation}

Perhaps we can also estimate the original amplitude of the oscillations $W_{\text{osc}}$. Let us assume that, when inflation ends, we enter into the stage of vector field oscillations very quickly. Then, we can consider
\begin{equation}\label{eq:osc_assump}
 (\rho_{A})_{\text{end}}\approx(\rho_{A})_{\text{osc}}
 \qquad\text{and}\qquad
 (\rho_{\sigma})_{\text{end}}\approx(\rho_{\sigma})_{\text{osc}}
 \quad\Longrightarrow\mathcal{R}_{\text{end}}\approx\mathcal{R}_{\text{osc}}\,.
\end{equation}
We can first obtain \mbox{$(\rho_{A})_{\text{end}}=\left(\mathcal{R}\rho_{\sigma}\right)_{\text{end}}$} where \mbox{$(\rho_{\sigma})_{\text{end}}\simeq V(\sigma_{\text{end}})$}. The total energy density of the vector field is given by a kinetic term plus a potential term \mbox{$\rho_A=\rho_{\textrm{kin}}+V_A$} where \mbox{$\rho_{\textrm{kin}}\equiv  a_2^{-2}f\dot{A}^2_z/2$} and \mbox{$V_A\equiv a_2^{-2}m^2A^2_z/2$}. Now for a heavy vector field that is oscillating we have \mbox{$\bar{\rho}_{\textrm{kin}}=\bar{V}_{A}$}, where the bar denotes the average, and therefore \mbox{$({\rho}_{A})_{\text{osc}}=a_2^{-2}m_{A}^{2}\bar{A^{2}}$}. Then assuming (for sinusoidal oscillations) \mbox{$\bar{A^{2}}\approx\hat{A}^{2}/2$}, where $\hat{A}$ is the amplitude of oscillations of the comoving vector field. We can then find
\begin{equation}
({\rho}_{A})_{\text{osc}}=\frac12h_{0}^{2}\phi_0^{2}W_{\text{osc}}^{2},
\end{equation}
where we have now neglected the residual anisotropic stress $\Sigma$ as it quickly vanishes during vector field oscillations \cite{Wagstaff} and we considered that \mbox{$W=A/a$} after the end of inflation. With the assumption in eq.~(\ref{eq:osc_assump}) we find
\begin{equation}\label{eq:W_osc}
 W_{\text{osc}}^{2}\approx
 \frac{2V(\sigma_{\text{end}})\mathcal{R}_{\text{end}}}{h_{0}^{2}\phi_0^2},
\end{equation}
then with eq.~(\ref{eq:g-general1}) and \mbox{$H_{*}\approx H_{\text{end}}$} we obtain
\begin{equation}\label{eq:g-general}
 \left|g\right|
 \approx\frac16\left(\frac{h_{0}\phi_0}{3n\pi m_P\zeta}\right)^{2}
 \frac{\mathcal{R}_{\text{dec}}^{2}}{\mathcal{R}_{\text{end}}}.
\end{equation}

If the inflaton does not decay too early, both fields will be coherently oscillating and \mbox{$\mathcal{R}=\text{constant}$}. The vector field can then decay into a matter dominated background where
\begin{equation}\label{eq:RdeDend2-1}
 n=1\qquad\text{and}\qquad
 \mathcal{R}_{\text{dec}}\simeq\mathcal{R}_{\text{end}}.
\end{equation}

However, for a better chance at generating statistical anisotropy, let us consider the case where the inflaton decays rapidly into relativistic particles, and the vector curvaton decays later into the radiation dominated Universe, hence
\begin{equation}\label{eq:RdeDend}
 n=\frac{4}{3}\qquad\text{and}\qquad\mathcal{R}_{\text{dec}}
 \simeq\varepsilon^{-1/2}\mathcal{R}_{\text{end}}.
\end{equation}

With this optimal scenario in eq.~(\ref{eq:RdeDend}), and using eq.~(\ref{eq:varepsilon-def}), the statistical anisotropy in eq.~(\ref{eq:g-general}) becomes
\begin{equation}\label{eq:g-general-R2}
 \left|g\right|
 \approx\sqrt{\frac23}
 \frac{\phi_0\mathcal{R}_{\text{end}}}{m_Ph_0\zeta|\Gamma_f(\sigma_*)|}.
\end{equation}

Therefore for Planck detectable statistical anisotropy \mbox{$\left|g\right|\gtrsim0.02$} \cite{planck}, in the best case scenario from eq.~(\ref{eq:g-general-R2}), an upper bound on the gauge coupling can be obtained
\begin{equation}\label{eq:h0_g-bound}
 h_{0}\lesssim 50
\sqrt{\frac{2}{3}}
 \frac{\phi_0\mathcal{R}_{\text{end}}}{m_{P}
 \zeta|\Gamma_{f}(\sigma_{*})|}.
\end{equation}

\subsection{Anisotropic non-Gaussianity}
The vector curvaton may also generate non-Gaussianity in the primordial curvature perturbation.\footnote{In this paper we discuss only the bispectrum, for the trispectrum see ref.~\cite{cesar}.} The latest CMB observations provide a hint for the detection of the non-linearity parameter $f_{\text{NL}}$ which characterizes non-Gaussianity \mbox{$f_{\text{NL}}^{local}=32\pm21$} (at 1$\sigma$) \cite{wmap}. For the spectra in eq.~(\ref{eq:spectra}) and using eq.~(\ref{eq:g}), from ref.~\cite{fnlanis} the non-linearity parameter was found to be
\begin{equation}
 \frac{6}{5}f_{\text{NL}}^{eq}= g^{2}\frac{3}{2\hat{\Omega}_{A}}\left(1-\frac{7}{8}\hat{W}_{\perp}^{2}\right) \qquad\text{and}\qquad
 \frac{6}{5}f_{\text{NL}}^{local}
 = g^{2}\frac{3}{2\hat{\Omega}_{A}}\left(1-\hat{W}_{\perp}^{2}\right),
\end{equation}
where \mbox{$\hat{W}_{\perp}\equiv|\mathbf{\hat{W}_{\perp}}|$}, with $\mathbf{\hat{W}_{\perp}}$ being the projection of the unit vector $\mathbf{\hat{W}}$ to the plane defined by the three $\mathbf{k}_{1,2,3}$ vectors which determine the bispectrum. Therefore the non-Gaussianity is itself anisotropic, it has directional dependence (given through $\hat{W}_{\perp}$) related to the preferred direction in the spectrum. Further more, the amplitude of non-Gaussianity is correlated to the amplitude of statistical anisotropy in the spectrum (given by $g$). These effects are signature predictions of a vector field contribution to the primordial curvature perturbation.

Using eq.~(\ref{eq:g-general}), the amplitude of non-Gaussianity generated in the sudden-symmetry breaking approximation in eq.~(\ref{eq:osc_assump}) is given by
\begin{equation}\label{eq:fNL_general}
 f_{\text{NL}}\simeq\frac{5}{4}g^{2}\frac{n}{\mathcal{R}_{\text{dec}}}
 \approx\frac{5}{144n^{3}}\left(\frac{h_{0}\phi_0}{3\pi m_P\zeta}\right)^{4}
 \frac{\mathcal{R}_{\text{dec}}^{3}}{\mathcal{R}_{\text{end}}^{2}}.
\end{equation}
In the best case scenario with eq.~(\ref{eq:RdeDend}), the amplitude of non-Gaussianity becomes
\begin{equation}\label{eq:fNL_general-R2}
 f_{\text{NL}}\simeq\frac{5}{3}g^{2}\frac{\sqrt{\varepsilon}}{\mathcal{R}_{\text{end}}}.
\end{equation}
Therefore for Planck detectable non-Gaussianity \mbox{$f_{\text{NL}}\gtrsim\mathcal{O}(1)$}, in the best case scenario, the following upper bound on the gauge coupling constant must be satisfied
\begin{equation}\label{eq:h0_f-bound}
 h_{0}
 \lesssim\frac{25\sqrt{6}}{3888}\frac{\mathcal{R}_{\text{end}}^{2}}{\pi^{2}
 |\Gamma_{f}^{3}(\sigma_{*})|}\left(\frac{\phi_0}{m_{P}\zeta}\right)^{5}.
\end{equation}

\section{Vector Scaling SUGRA Hybrid Inflation}\label{SUSY_Hybrid}


We now embed the above vector curvaton into a well motivated model of SUSY GUT Hybrid Inflation \cite{susyhybrid,sugrahybrid}. As described in previous sections, through the vector-scalar coupling, we will obtain a model of vector scaling Hybrid Inflation.

Consider a simple SUSY GUT model based on the gauge group \mbox{$G=G_{\rm SM}\times U(1)_{B-L}$} where $G_{\rm SM}$ is the Standard Model gauge group. This model naturally incorporates \cite{susyhybrid,sugrahybrid} the standard SUSY realisation of Hybrid Inflation. The gauge group $G$ can arise from concrete GUT models with larger gauge symmetry. For example the "semi-shifted" \cite{Lazarides:2008nx} inflationary scenario, which is based on the extended supersymmetric Pati-Salam model, a $U(1)_{B-L}$ remains unbroken during inflation which is then spontaneously broken at the end of inflation.

Consider a conjugate pair of $G_{\rm SM}$ singlet left handed superfields $\Phi$ and $\bar{{\Phi}}$ with charges $+1$ and $-1$ respectively with respect to the $U(1)_{B-L}$ gauge symmetry which thus break the \mbox{$U(1)_{B-L}$} by their VEVs. Consider also a gauge singlet left handed superfield $S$ which will act as our inflaton. The flatness of the inflationary trajectory is normally guaranteed by a $U(1)$ R-symmetry on $S$ (which only allows terms linear in $S$ in the superpotential $W$ to all orders), however in our model we will consider a discrete $Z_n$ symmetry under which \mbox{$S\rightarrow Se^{2\pi i/n}$} and \mbox{$W\rightarrow We^{2\pi i/n}$}. Note, in passing, that global continuous symmetries such as the $U(1)$ R-symmetry normally considered can effectively arise \cite{Lazarides:1985bj} from the rich discrete symmetry groups encountered in many compactified string theories (see e.g. ref.~\cite{classifications}). Allowing for non-renormalisable terms, the most general form of the superpotential relevant to inflation and permitted by the symmetries is
\begin{equation}
 W= S\sum_{k_1,k_2=0}^{\infty}A_{k_1k_2}\left(\Phi\bar{{\Phi}}\right)^{k_1}
 \left(S^n\right)^{k_2},
\end{equation}
where \mbox{$A_{k_1k_2}$} are coefficients with varying dimensions (for a similar analysis see ref.~\cite{Lazarides:1998zf}). Expanding the superpotential we find that for $n\geq3$
\begin{equation}
 W=\kappa S\left(\Phi\bar{{\Phi}}-M^{2}\right) + \text{"non-renormalisable terms"},
\end{equation}
with \mbox{$A_{00}=-\kappa M^2$} and \mbox{$A_{10}=\kappa$} and the parameters $\kappa$ and $M$ are made positive by field redefinitions. We will take the view that terms in the superpotential with $n\neq0$ are $m_P$ suppressed corrections to the renormalisable superpotential given above, for example the coefficient \mbox{$A_{01}=-c\kappa M^2/m_P^n$} for some expansion coefficient $c$ of order unity.

Through the remainder of this paper, unless stated otherwise, we will associate the scale \mbox{$M\simeq M_{\rm GUT}$}, where \mbox{$M_{\rm GUT}=2.86\times10^{16}$}GeV is the scale of grand unification, so that the spontaneous symmetry breaking corresponds to the breaking of grand unification. The SUSY minimum is at $\langle S\rangle=0$ and \mbox{$\langle \Phi\rangle=\langle \bar{\Phi}\rangle=M$}.

Models in supergravity are defined by three fundamental functions; the superpotential $W$, the K\"{a}hler potential $K$ and the gauge kinetic functions $f_{ab}$. The scalar potential in supergravity has the form
\begin{equation}\label{spot_SUGRA}
V=e^{K/m_P^2}\left[F_{\Phi_i}K^{-1}_{ij*}F_{\Phi^*_j} -3\frac{|W|^2}{m_P^2} \right]
+\frac12 \sum_{a,b}\left[\text{Re}f_{ab}(\Phi_i)\right]^{-1}h_ah_bD_aD_b
\end{equation}
where
\begin{equation}
K_{ij*}=\frac{\partial^2K}{\partial\Phi_i\partial\Phi^*_j}\,,\quad F_{\Phi_i}=\frac{\partial W}{\partial \Phi_i}+\frac{W}{m_P^2}\frac{\partial K}{\partial \Phi_i} \quad\text{and}\quad
D_a=\Phi_i(T_a)^i_j\frac{\partial K}{\partial \Phi_j}+\xi_a.
\end{equation}
The subscripts \mbox{$a,b,\cdots$} label the generators $T_a$ of the gauge group with gauge couplings $h_a$. The $\xi_a$ are Fayet-Iliopoulos D-terms, which can only exist for $U(1)$ gauge groups. In our model, only the gauge kinetic function associated with the $U(1)_{B-L}$, which we simply write as $f$, will not equal to unity. All kinetic functions associated with other gauge symmetries will be set equal to one. And for our model we have \mbox{$\Phi_i=(S,\Phi,\bar{{\Phi}})$}. The D-flatness is given by \mbox{$|\Phi|=|\bar{\Phi}|$} i.e. \mbox{$\bar{\Phi}^*=\Phi e^{i\theta}$} where we choose $\theta=0$ so that the SUSY vacua are contained. Bringing the fields $\Phi$ and $\bar{{\Phi}}$ on the real axis by appropriate $U(1)_{B-L}$ transformations, we write \mbox{$\Phi=\bar{\Phi}\equiv\phi/2$} where $\phi$ is a normalised real scalar field. It will also be useful to define another real scalar field $\sigma$ where \mbox{$|\sigma|\equiv\sqrt{2}|S|$} (see below). The supersymmetric minimum is then found at \mbox{$\sigma=\sigma_0=0$} and \mbox{$\phi=\phi_0=\pm2M$}.

The effective mass-squared $m_{\phi}^{2}$ for the field \mbox{$\phi$} is given
by\footnote{If the waterfall field is coupled to the gauge field as well, as we
assume in this paper, there is also a contribution from the gauge field to its
mass-squared. According to eq.~(\ref{model}), this contribution is
$\sim (h_0 W)^2$, where \mbox{$W=|{\bf W}|$} with \mbox{$W_i=\sqrt f A_i/a$}
and \mbox{$f\equiv 1$} at the end of inflation. Assuming that $W$ is no larger
than the value of the physical vector field condensate after the end of
inflation, then \mbox{$\rho_W<\rho_{\rm inf}=\kappa^2 M^4$}, suggests that
\mbox{$W<\kappa M/h_0$}, where \mbox{$\rho_W\sim (h_0 M W)^2$} is the density
of the vector field after the phase transition which ends inflation. This
condition guarantees that the contribution of the vector field to the tachyonic
mass of the waterfall is subdominant. However, perturbations of the vector
field can perturb the phase transition and generate a contribution to $\zeta$
as shown in ref.~\cite{soda}. We do not consider this mechanism in this paper.}
\begin{equation}
 m_{\phi}^{2}\simeq\kappa^{2}\left(|S|^{2}-M^{2}\right),
\end{equation}
hence the field becomes tachyonic at the critical value \mbox{$\sigma=\sigma_{c}=\pm\sqrt{2}M$}. This is the field value at which the waterfall takes effect. For \mbox{$|\sigma|>|\sigma_c|$} the potential is minimised at \mbox{$\Phi=\bar{\Phi}=0$} (i.e. \mbox{$\phi=0$}) and inflation is driven by the false vacuum energy density \mbox{$\kappa^2M^4$}. Inflation is only stable for values \mbox{$|\sigma|>|\sigma_{c}|$}, we can therefore assume that inflation ends abruptly at the waterfall.

The K\"{a}hler potential is a real function of the superfields and must be invariant under the symmetries of the theory. In our model it is therefore a real function of \mbox{$|S|^2,|\Phi|^2,|\bar{\Phi}|^2,\Phi\bar{\Phi}$} and $S^n$. We expect the K\"{a}hler potential to be an $m_P$ suppressed expansion of the superfields. On the inflationary trajectory where \mbox{$\Phi=\bar{\Phi}=0$}, it is clear that the matrix \mbox{$K_{ij*}$} becomes diagonal. We also notice that the terms \mbox{$F_{\Phi}=F_{\bar{\Phi}}=0$} on the trajectory, therefore the only terms in the K\"{a}hler potential that will contribute on the trajectory are
\begin{equation}\label{eq:Kahler}
 K= \sum_{k_1,k_2=0}^{\infty} \frac{|S|^{2k_1}}{m_P^{2k_1+nk_2-2}} \left[a_{k_1k_2}\left(S^n\right)^{k_2}+h.c\right].
\end{equation}
The \mbox{$a_{k_1k_2}$} are dimensionless coefficients of order one. Hence \mbox{$K=|S|^2-(\alpha/4)|S|^4/m_P^2+\cdots$}, with \mbox{$a_{00}=0$}, \mbox{$a_{10}=1/2$} and \mbox{$a_{20}=-\alpha/8$} where \mbox{$|\alpha|\sim1$} is a real parameter.

The scalar potential, given in eq.~(\ref{spot_SUGRA}), on the inflationary trajectory can then be parameterized by
\begin{equation}\label{eq:spot}
 V= \kappa^2M^4\sum_{k_1,k_2=0}^{\infty} P_{k_1k_2}\frac{|S|^{2k_1}(S^n)^{k_2}}{m_P^{2k_1+nk_2}}+h.c.
\end{equation}
where the \mbox{$P_{k_1k_2}$} are dimensionless coefficients that are functions of the superpotential coefficients \mbox{$A_{k_1k_2}$} and the K\"{a}hler potential coefficients \mbox{$a_{k_1k_2}$}. Switching to polar coordinates \mbox{$S=|S|e^{i\vartheta}=|\sigma|e^{i\vartheta}/\sqrt2$}, the scalar potential, calculated up to terms suppressed by $m_P^4$, is given by
\begin{equation}\label{eq:potential}
 V(\sigma,\phi=0)\simeq\kappa^{2}M^{4}\left\{1+\frac{\alpha}{2}\left(\frac{\sigma}{m_{P}}\right)^{2}
 +\beta\left(\frac{\sigma}{m_{P}}\right)^{4}+
 2\gamma(n+1)\left(\frac{|\sigma|}{\sqrt2m_P}\right)^n\cos n\vartheta
 +\cdots\right\},
\end{equation}
where \mbox{$8\beta=1+7\alpha/2+2\alpha^2-18(a_{30}+c.c)$} and \mbox{$\gamma=c+a_{01}-a_{11}$} and we assumed $\gamma$ to be real. The phase $\vartheta$ will be homogenised by inflation and will assume a constant value. Minimising the potential with respect to the phase we find that for \mbox{$\gamma<0$}, \mbox{$n\vartheta=2\pi k$} for integer $k$. Then applying $l$ times the generators of the $Z_n$ symmetry we obtain \mbox{$n\vartheta=2\pi k+2\pi l$}, and appropriately choosing $l$ we can set \mbox{$\vartheta=0$}. For \mbox{$\gamma>0$}, the potential is minimised with \mbox{$n\vartheta=(2k+1)\pi$}. Then applying $l$ times the generators of the $Z_n$ symmetry we obtain \mbox{$n\vartheta=(2k+1)\pi+2\pi l$}, and again appropriately choosing $l$ we can set \mbox{$\vartheta=\pi/n$}.

Looking at the potential in eq.~(\ref{eq:potential}) we observe that the extra terms arising from the $Z_n$ symmetry (as opposed to the continuous R-symmetry normally considered) are substantially subdominant for \mbox{$n\geq3$}. This potential reduces to the known potential for the canonical K\"{a}hler potential case where \mbox{$\alpha=0$}.

The one-loop radiative corrections to the scalar potential on the inflationary trajectory which are calculated by using the Coleman-Weinberg formula \cite{Coleman:1973jx} are given by
\begin{equation}
 \Delta V_{\rm 1-loop}=\frac{\left(\kappa M\right)^{4}\mathcal{N}}{32\pi^{2}}
 \left(2\ln\frac{\kappa^{2}\sigma^{2}}{2Q^{2}}+f_c(x)\right),
\end{equation}
with \mbox{$f_c(x)\equiv (x+1)^2\ln (1+1/x)+(x-1)^2\ln (1-1/x)$} where \mbox{$x\equiv \sigma^2/2M^2$} and \mbox{$\mathcal{N}$} is the dimensionality of the representation to which \mbox{$\Phi$} and \mbox{$\bar{{\Phi}}$} belong (in this model \mbox{$\mathcal{N}=1$}), and $Q$ is some renormalisation scale (we note that the slope will not depend on this scale). In standard SUSY Hybrid Inflation, these radiative corrections are necessary to generate a slope on the inflationary valley \cite{susyhybrid} when considering the minimal K\"{a}hler potential.

The non-minimal K\"{a}hler potential above induces non-canonical kinetic terms for the scalar fields. However, our dynamical analysis discussed in the previous section which leads to the VSSR attractor \cite{Wagstaff}, assumes canonically normalised fields. Therefore we have to consider sub-Planckian field values \mbox{$\sigma< m_{P}$} when the backreaction takes effect so that the fields are approximately canonically normalised and we can trust our dynamical analysis.

In the inflationary valley at \mbox{$\phi=0$}, the inflaton \mbox{$\sigma$} obtains a contribution to its mass squared from the non-canonical K\"{a}hler potential, given by \mbox{$V''(\phi=0)\simeq3\alpha H^{2}$} (i.e. \mbox{$\eta\simeq\alpha$}). Therefore the field would normally be fast-rolling and the valley could not support standard slow-roll inflation unless the coefficient \mbox{$\alpha$} is suppressed. However, considering the VSSR attractor, there is the possibility to obtain slow-roll inflation even without fine-tuning the non-canonical coefficient $\alpha$.

The dimensionless model parameter in eq.~(\ref{eq:mod-parameters}) for the scalar potential given in eq.~(\ref{eq:potential}) is
\begin{equation}\label{eq:lambda0}
 \lambda_{0}\simeq\sqrt{\frac{3}{2}}\left[\alpha\left(\frac{\sigma}{m_{P}}\right)
 +\left(4\beta-\frac{1}{2}\alpha^{2}\right)\left(\frac{\sigma}{m_{P}}\right)^{3}
 +\frac{\kappa^{2}}{8\pi^{2}}\left(\frac{m_{P}}{\sigma}\right)+\cdots\right].
\end{equation}
For \mbox{$\sigma\ll m_{P}$} there is a competition between the first and last term in the expression above depending on the value of the relevant parameters. For large coupling constant \mbox{$\kappa\sim0.1$} and \mbox{$|\alpha|\sim1$} there is the possibility that the radiative correction term comes into play near the end of inflation. For this term to be important during inflation fine-tuning of the parameter \mbox{$\alpha$} is required. This is usually considered as to obtain a red spectrum of inflaton perturbations. However, in this paper, we will consider the natural value \mbox{$|\alpha|\sim1$} and show that a red spectrum can still be obtained if the cosmological scales exit during the VSSR attractor.

Let us assume that the first term in eq.~(\ref{eq:lambda0}) dominates over the radiative corrections until the end of inflation at \mbox{$\sigma=\sigma_c$}. The condition for this to occur is given by
\begin{equation}\label{eq:radiative-cond}
 \kappa<4\pi\sqrt{|\alpha|}\left(\frac{M}{m_{P}}\right).
\end{equation}
For \mbox{$|\alpha|=1$} we find the upper bound \mbox{$\kappa\lesssim0.148$}. We will later verify that this condition is satisfied for a specific model example. As long as this condition is satisfied we can approximate the dimensionless model parameter in eq.~(\ref{eq:lambda0}) as
\begin{equation}\label{eq:lambda0_1st}
 \lambda_{0}\simeq\alpha\sqrt{\frac{3}{2}}\left(\frac{\sigma}{m_{P}}\right).
\end{equation}
We note that in eq.~(\ref{eq:lambda0_1st}), if \mbox{$|\alpha|\sim1$} we find \mbox{$|\lambda_{0}|\lesssim1$}. It will also be useful to find the first derivative of the dimensionless model parameter
\begin{equation}\label{eq:lambda'}
 \sqrt{\frac{2}{3}}m_{P}\left|\lambda_{0}'\right|
 =\left|\eta-2\epsilon\right|\simeq
 \Big|\alpha +3\left(4\beta-\frac{1}{2}\alpha^{2}\right)\left(\frac{\sigma}{m_{P}}\right)^{2}+\cdots \Big|.
\end{equation}

\section{An exponential gauge kinetic function}

The form of the gauge kinetic function is constrained by its holomorphicity and the symmetries of the model. The $Z_n$ symmetry prevents the appearance of terms in $f$ which are linear in $S$, but combinations $S^n$ are allowed. Combinations $\Phi\bar{\Phi}$ are also allowed by the symmetries, but these will not contribute on the inflationary trajectory. Consider a gauge field with an exponential gauge kinetic function of the following form
\begin{equation}\label{eq:gauge_fn1}
 f(S^n)=\exp\left[q\left(\frac SM\right)^n\right]
 =\exp\left[q\left(\frac{|\sigma|}{\sqrt2M}\right)^ne^{in\vartheta}\right],
\end{equation}
where the vector field becomes canonically normalised with \mbox{$f\rightarrow1$} as the inflaton settles into the SUSY vacuum \mbox{$\sigma_0=0$} at the end of inflation. For \mbox{$\gamma<0$}, where we can set \mbox{$\vartheta=0$} we choose $q$ to be a real positive parameter and for \mbox{$\gamma>0$} where we can set \mbox{$\vartheta=\pi/n$} we choose $q$ to be a real negative parameter so that the exponent is always positive
\begin{equation}\label{eq:gauge_fn1a}
 f(\sigma)=e^{|q|\left(|\sigma|/\sqrt 2M\right)^n}.
\end{equation}
The dimensionless model parameter defined in eq.~(\ref{eq:mod-parameters}) is then given by
\begin{equation}\label{eq:gauge_fn1b}
 \Gamma_{f}(\sigma)=|q|n{\frac{\sqrt 3}{2}}\left(\frac{m_{P}}{M}\right)\left(\frac{\sigma}{|\sigma|}\right)
 \left(\frac{|\sigma|}{\sqrt 2M}\right)^{n-1}.
\end{equation}
For this model, the dimensionless model parameter $\Gamma_f$ is not a constant for \mbox{$n\neq1$}. The existence condition \cite{Wagstaff} for the VSSR attractor demands that $\Gamma_f$ and $\lambda_0$ have the same sign, hence we require that the K\"{a}hler potential parameter $\alpha>0$. We now choose \mbox{$\sigma>0$} for definiteness without loss of generality.

We must now identify the conditions on the model parameters which will lead to the VSSR attractor, namely conditions in eqs.~(\ref{eq:conditions_1-2-3}) and (\ref{eq:CondA-D}). For \mbox{$|q|\sim1$}, \mbox{$\sigma>\sigma_c$} and \mbox{$n\geq3$} we find \mbox{$\Gamma_{f}\gg1$}, hence condition I in eq.~(\ref{eq:conditions_1-2-3}) is readily satisfied. Considering the approximation in eq.~(\ref{eq:lambda0_1st}) with \mbox{$\alpha\sim1$} we have \mbox{$\lambda_{0}\lesssim1$}, therefore condition II in eq.~(\ref{eq:conditions_1-2-3}) is also readily satisfied. Then requiring that condition III in eq.~(\ref{eq:conditions_1-2-3}) is satisfied right up until the end of inflation at \mbox{$\sigma=\sigma_{c}$} we obtain the following condition on the model parameters
\begin{equation}\label{eq:bound1}
 |q|\alpha n>4.
\end{equation}

In this model the VSSR attractor is a moving point in phase-space (because of \mbox{$\lambda_0=\lambda_0(\sigma)$} and \mbox{$\Gamma_f=\Gamma_f(\sigma)$}), therefore we have the additional conditions in eq.~(\ref{eq:CondA-D}) to ensure that solutions are able to reach the moving attractor and be dragged along with it. For this model, in which \mbox{$\Gamma_{f}'/\Gamma_{f}=(n-1)/\sigma$} and \mbox{$\lambda_{0}'/\lambda_{0}=1/\sigma$}, we find from eqs.~(\ref{eq:CondA-1})-(\ref{eq:CondD-1}) that the two tightest conditions are \mbox{$A,C<1$}, i.e.
\begin{equation}\label{eq:conditions_A-B}
 A=4\sqrt{\frac{2}{3}}m_{P}\frac{\left|\Gamma_{f}'\right|}{\Gamma_{f}^{2}}\lesssim1
 \qquad\text{and}\qquad
 C=4\sqrt{\frac{2}{3}}m_{P}\left|\frac{\Gamma_{f}'}{\Gamma_{f}^{2}}
 -\frac{\lambda_{0}'\Gamma_{f}
 +\lambda_{0}\Gamma_{f}'}{\Gamma_{f}\left(\lambda_{0}\Gamma_{f}-6\right)}\right|\lesssim1.
\end{equation}
For $A<1$ to be satisfied up until the end of inflation we require that \mbox{$|q|\gtrsim8(n-1)/3n$} which can be easily satisfied even for large $n$ giving \mbox{$|q|\gtrsim8/3$}. Then for condition $C<1$ to be satisfied up until the end of inflation, and assuming condition $A<1$ is satisfied, we obtain the following bound on the parameters
\begin{equation}\label{eq:bound2}
|q|\alpha n-4\gtrsim\frac{8}{3}\alpha n.
\end{equation}
The condition $C<1$ simply tells us that we cannot be too close to the bifurcation point at \mbox{$\lambda_{0}\Gamma_{f}=6$} (see ref. \cite{Wagstaff}) as this is where the VSSR attractor disappears and we enter into a stage of fast-roll inflation. The bound in eq.~(\ref{eq:bound2}) can be written as \mbox{$|q|\gtrsim8/3+4/\alpha n$} which for large $n$ goes to the same limit as that of condition $A<1$ i.e. \mbox{$|q|\gtrsim8/3$}, but otherwise it is a stronger bound on parameter $q$. The bound in eq.~(\ref{eq:bound2}) is also stronger than the bound in eq.~(\ref{eq:bound1}) and therefore guarantees that inflation ends before condition III is violated. This bound should be made strong to guarantee that the system remains at the VSSR attractor right up until the end of inflation. For a given $q$ and $n$, the bound on $\alpha$ becomes
\begin{equation}\label{eq:alpha-bound}
 \alpha\gtrsim 4\left[n\left(|q|-\frac83\right)\right]^{-1}.
\end{equation}

\subsection{Properties of vector scaling slow-roll inflation}

We will now assume that the VSSR attractor is obtained. From eq.~(\ref{eq:efoldings}), the e-foldings $N_{\text{att}}$ generated during VSSR attractor is given by
\begin{equation}
 N_{\text{att}}=\frac{|q|}{4}\left[\left(\frac{\sigma_i}{\sqrt 2M}\right)^n-1\right]
 \quad\Longrightarrow\quad
 \frac{\sigma_{i}}{m_{P}}=\frac{\sqrt 2M}{m_{P}}\left(\frac{4 N_{\text{att}}}{|q|}+1\right)^{1/n},
\end{equation}
where $\sigma_i$ denotes the field value at the start of the attractor and the field value at the end of the attractor is given by \mbox{$\sigma_{\text{end}}=\sigma_c$}. Requiring sub-Planckian initial field values \mbox{$\sigma_i<m_P$}, an upper bound on the total number of e-folds is obtained for a given $q$ and $n$. For \mbox{$|q|=3$} and \mbox{$n=3$} we find that \mbox{$N_{\text{att}}\lesssim10^5$} which is many more e-foldings than the number needed to solve the horizon and flatness problems. As an example, evaluating the above for prompt reheating where \mbox{$N_{\text{att}}\simeq67$} to solve the cosmological problems, with \mbox{$n=3$} and \mbox{$|q|=5$} we obtain \mbox{$\sigma_i/m_P\simeq0.06$} and therefore we are well into the sub-Planckian regime and the assumption that the attractor solution is obtained by this field value is justified. Similarly, the field value when cosmological scales (in particular the pivot scale) exit the horizon at \mbox{$N_{*}\simeq60$} is given by
\begin{equation}\label{eq:N_k}
 \frac{\sigma_{*}}{\sqrt2M}=\left(\frac{4 N_{*}}{|q|}+1\right)^{1/n}.
\end{equation}

For the gauge kinetic function under consideration, we find from eq.~(\ref{eq:V_eff}) the reduced effective scalar potential slope \mbox{$V'_{\textrm{eff}}/V'\simeq(4/|q|\alpha n)(\sqrt 2M/\sigma)^n\ll1$}. The slow-roll parameters given in eq.~(\ref{eq:slow-roll_gen}) become
\begin{equation}
 \epsilon_{\textrm{H}}\simeq\frac{2\alpha}{|q|n}\left(\frac{\sigma}{m_{P}}\right)^2\left(\frac{\sqrt 2M}{\sigma}\right)^n
 \quad\text{and}\quad
 \eta_{\textrm{H}}\simeq\epsilon_{\textrm{H}}+\frac{2(2-n)}{|q|n}\left(\frac{\sqrt 2M}{\sigma}\right)^n,
\end{equation}
hence \mbox{$\epsilon_{\textrm{H}},\eta_{\textrm{H}}\ll1$} for \mbox{$n\geq3$} and \mbox{$m_P>\sigma>\sigma_c=\sqrt2M$}.

The attractor solution for the vector-to-scalar energy density ratio $\mathcal{R}$, given in eq.~(\ref{eq:RandSigma}), for this model becomes
\begin{equation}
 \mathcal{R}(\sigma)\simeq
 \frac{2}{(qn)^{2}}\left(\frac{M}{m_{P}}\right)^{2}\left(\frac{\sigma}{\sqrt 2M}\right)^{2(1-n)}\left[|q|\alpha n\left(\frac{\sigma}{\sqrt 2M}\right)^n-4\right].
\end{equation}
Thus, we see that for \mbox{$n\geq3$}, $\cal R$ is increasing with time. Therefore, we must have $\mathcal{R}\ll1$ at the end of inflation so that the vector field remains safely subdominant throughout VSSR inflation. The value of ${\cal R}(\sigma)$ at the end of inflation is
\begin{equation}\label{eq:Rend}
 \mathcal{R}_{\text{end}}\simeq
 2\left(\frac{M}{m_{P}}\right)^{2}\left[\frac{|q|\alpha n-4}{(qn)^{2}}\right]
 \gtrsim\frac{16}{3}\left(\frac{M}{m_P}\right)^2\frac{\alpha}{q^2n}
 \gtrsim\frac{64}{3}\left(\frac{M}{m_P}\right)^2\left[(qn)^2\left(|q|-\frac83\right)\right]^{-1},
\end{equation}
where we used eqs.~(\ref{eq:bound2}) and (\ref{eq:alpha-bound}).
With \mbox{$n=3$}, \mbox{$|q|=3$} and \mbox{$\alpha=4$} which saturates the bound in eq.~(\ref{eq:alpha-bound}), we find that \mbox{$\mathcal{R}_{\text{end}}\simeq 1\times10^{-4}$}. For large $n$ the ratio decreases as \mbox{$\mathcal{R}_{\text{end}}\propto1/n$}. 

\subsection{The primordial curvature perturbation and gravitational waves}

From eqs.~(\ref{eq:dp/p}) and (\ref{eq:N_k}), to leading order in the potential in eq.~(\ref{eq:potential}) i.e. \mbox{$V(\sigma_*)\simeq\kappa^{2}M^{4}$}, we obtain
\begin{equation}\label{eq:dp/p1}
 \frac{2}{5}\zeta\simeq
 \frac{\kappa |q|n}{20\sqrt{6}\pi}\left(\frac{M}{m_{P}}\right)\left(\frac{4N_*}{|q|}+1\right)^{(n-1)/n},
\end{equation}
hence the COBE normalisation with \mbox{$M=M_{\rm GUT}$}, \mbox{$N_*=60$}, \mbox{$n\geq3$}, and \mbox{$|q|\gtrsim3$} leads to the upper bound on the coupling constant \mbox{$\kappa \lesssim1.5\times10^{-3}$}. This bound for the coupling constant $\kappa$ readily satisfies the condition in eq.~(\ref{eq:radiative-cond}) for \mbox{$\alpha=4$} i.e. \mbox{$\kappa\lesssim0.30$}. Therefore the approximation used in eq.~(\ref{eq:lambda0_1st}), where the radiative corrections to the potential remain negligible until the end of inflation, is well justified. The COBE normalisation is satisfied for exactly \mbox{$M=M_{\rm GUT}$} whereas in the standard SUSY Hybrid Inflation model \cite{susyhybrid,sugrahybrid} the maximal $M$ which can be achieved is somewhat below the SUSY GUT VEV. This is an additional attractive feature of our model as the spontaneous symmetry breaking which ends inflation corresponds exactly to the breaking of grand unification.

We observe that the second term in the square brackets in eq.~(\ref{eq: spectral index}) for the spectral index is positive, therefore a red spectrum of perturbations is obtained. For this model, using eqs.~(\ref{eq:lambda0_1st}), (\ref{eq:gauge_fn1}) and (\ref{eq:N_k}), we find the following result for the spectral index
\begin{equation}
 n_{s}\simeq1-\frac{8}{|q|n}\left(\frac{4N_{*}}{|q|}+1\right)^{-1}
 \left[\alpha\left(\frac{M}{m_{P}}\right)^{2}\left(\frac{4N_{*}}{|q|}+1\right)^{2/n}+(n-1)\right].
\end{equation}
For \mbox{$n\geq3$} and \mbox{$N_*\simeq60$} we observe that the spectral index depends very weakly on the model parameters \mbox{$q$} and $\alpha$ for \mbox{$|q|,\alpha\sim1$}. In fact the result above is well approximated by the
compact formula
\begin{equation}
 n_{s}\simeq1-\frac{2(n-1)}{nN_*}.
\end{equation}
We find that for \mbox{$n=3$} the spectral index is \mbox{$n_s\simeq0.978$} and for \mbox{$n\gg1$} we obtain \mbox{$n_s\simeq0.967$}, which fit very well within the ($1\sigma$) CMB observational bounds from WMAP7 on the spectral index at \mbox{$0.953\leq n_s\leq0.981$} \cite{wmap}. For large enough $n$, we can therefore obtain the observationally preferred central value for the spectral index \mbox{$n_s=0.967$}.

The running of the spectral index, given in eq.~(\ref{eq:running}), for this model becomes
\begin{equation}
 n_{s}'\simeq\frac{24}{\Gamma_{f}^{2}}\left[\eta-2\epsilon-2n(n-1)\left(\frac{m_P}{\sigma_*}\right)^2\right].
\end{equation}
Using eq.~(\ref{eq:N_k}) and with \mbox{$\epsilon\simeq(\alpha^2/2)(\sigma_*/m_P)^2$} and \mbox{$\eta\simeq\alpha$} for \mbox{$\alpha\sim1$} and \mbox{$n\geq3$} the running is well approximated by
\begin{equation}
 n_{s}'\simeq-\frac{2(n-1)}{nN_*^2}.
\end{equation}
We find that for \mbox{$n=3$} the running of the spectral index is \mbox{$n_s'\simeq-3.6\times10^{-4}$} and for \mbox{$n\gg1$} we obtain \mbox{$n_s'\simeq-5.4\times10^{-4}$}, which satisfies the CMB WMAP7 observational constraints on the running of the spectral index with no gravitational waves at \mbox{$-0.084 < n_s' < 0.020$} (at \mbox{$95\%\text{cf}$}) \cite{wmap}.

The tensor-to-scalar ratio, given in eq.~(\ref{eq:Tensor-scalar}) and using eq.~(\ref{eq:N_k}), for this model becomes
\begin{equation}
 r\simeq\frac{256}{(qn)^{2}}\left(\frac{M}{m_{P}}\right)^{2}\left(\frac{4N_*}{|q|}+1\right)^{2(1-n)/n}.
\end{equation}
With \mbox{$N_*=60$}, \mbox{$n\geq3$} and \mbox{$|q|\gtrsim3$} we find that the running is \mbox{$r\lesssim1\times10^{-6}$}, which satisfies the current CMB observational bound from WMAP7 on gravitational waves with no running at \mbox{$r < 0.36$} (at \mbox{$95\%\text{cf}$}) \cite{wmap} but is probably too small to ever be observable.

\subsection{Statistical anisotropy and anisotropic non-Gaussianity}

We will now consider the vector field to act as a vector curvaton. A vector field contribution to the curvature perturbation will induce statistical anisotropy and anisotropic non-Gaussianity \cite{stanis,fnlanis}. For the gravitational effect of the vector curvaton not to be suppressed we require that \mbox{$\varepsilon\lesssim1$} where $\varepsilon$ is defined in eq.~(\ref{eq:varepsilon-def}). From eq.~(\ref{eq:h0_bound1}) and using eq.~(\ref{eq:N_k}), the following upper bound on the gauge coupling constant may be obtained
\begin{equation}\label{eq:h_0}
 h_{0}^3\lesssim\frac{32\sqrt2\pi^{2}\zeta}{|q|n}
 \left(\frac{4N_*}{|q|}+1\right)^{(1-n)/n}.
\end{equation}
With \mbox{$N_*=60$}, \mbox{$n\geq3$} and \mbox{$|q|\gtrsim3$} the upper bound on the gauge coupling becomes \mbox{$h_0\lesssim 0.05$}. The gauge coupling for SUSY GUT is around \mbox{$h_0\sim0.7$} and therefore we expect the gravitational effect of the vector curvaton to be suppressed somewhat.

The statistical anisotropy induced in the spectrum, given in eq.~(\ref{eq:g-general-R2}) and using eq.~(\ref{eq:N_k}), for this model becomes
\begin{equation}
 \left|g\right|\approx\frac{4\sqrt2\mathcal{R}_{\text{end}}}{3|q|nh_0\zeta}
 \left(\frac{M}{m_P}\right)^{2}\left(\frac{4N_*}{|q|}+1\right)^{(1-n)/n},
\end{equation}
where $\mathcal{R}_{\text{end}}$ is given in eq.~(\ref{eq:Rend}). In eq.~(\ref{eq:h0_g-bound}) a bound on the gauge coupling is given for Planck \cite{planck} detectable statistical anisotropy. In this model the bound becomes \mbox{$h_0\lesssim1\times10^{-4}$} for parameter values \mbox{$\alpha\simeq4$} with \mbox{$N_*=60$}, \mbox{$n\geq3$} and \mbox{$|q|\gtrsim3$}. This coupling is far smaller than the SUSY GUT gauge coupling at \mbox{$h_0\sim0.7$} and therefore observable statistical anisotropy through the vector curvaton mechanism is not possible in this model unless the gauge coupling is highly fine-tuned.

The amplitude of non-Gaussianity is given by (see eq.~(\ref{eq:fNL_general-R2}))
\begin{equation}
 f_{\text{NL}}\simeq\frac53g^2\frac{\sqrt{\varepsilon}}{\mathcal{R}_{\text{end}}},
\end{equation}
where $g$ is given above, $\varepsilon$ in eq.~(\ref{eq:varepsilon-def}) and $\mathcal{R}_{\text{end}}$ in eq.~(\ref{eq:Rend}). For detectable non-Gaussianity \mbox{$f_{\text{NL}}\gtrsim\mathcal{O}(1)$} we have an additional bound on the gauge coupling given in eq.~(\ref{eq:h0_f-bound}). In this model the bound becomes \mbox{$h_0\lesssim3\times10^{-10}$} for parameter values \mbox{$\alpha\simeq4$} with \mbox{$N_*=60$}, \mbox{$n\geq3$} and \mbox{$|q|\gtrsim3$}. Clearly this gauge coupling is far too small and therefore observable non-Gaussianity form the vector curvaton mechanism is very unlikely in this model.\footnote{However, the contribution to statistical anisotropy in the spectrum and bispectrum from the end-of-inflation mechanism may be more substantial \cite{soda}.}

\section{Conclusions}

We have demonstrated that, in an inflationary model, when the inflaton also
modulates the kinetic function of a vector boson field, the backreaction to
the inflaton's variation is such that allows steep inflation despite sizable
K\"{a}hler corrections to the scalar potential and also produces a mildly red
spectral index of inflaton perturbations. In that respect it eliminates the
$\eta$-problem of inflation in supergravity (SUGRA).

We have applied the above to a model of SUGRA Hybrid Inflation, where the
waterfall field is taken to be the Higgs field of a Grand Unified Theory (GUT).
The vector field is taken to be one of the supermassive GUT bosons, which
becomes massive at the GUT phase transition that terminates inflation. The
gauge kinetic function of this field depends on the inflaton and is therefore
modulated during inflation. We have
shown that slow-roll inflation can take place with a generic K\"{a}hler
potential, that includes higher order corrections beyond the minimal form,
despite the fact that \mbox{$\eta={\cal O}(1)$}. Moreover, we have shown that
a red spectrum of perturbations is attained, in excellent agreement with
observations. Indeed, assuming an exponential
gauge kinetic function, we have shown that one can obtain for the spectral
index \mbox{$n_s\simeq 0.97-0.98$} with \mbox{$\eta\sim 1$}, with
negligible running and tensor fraction.

It is interesting to note that, in SUSY models based on
\mbox{$G_{\rm SM}\times U(1)_{B-L}$} such as ours, the baryon
asymmetry of the Universe can be easily generated via a
primordial non-thermal leptogenesis \cite{origin}, which takes
place during the direct inflaton decay to light particles as
shown in ref.~\cite{direct}. The cold dark matter in the
Universe can, in principle, consist of the lightest neutralino
as in many SUSY theories.

In principle the vector field can also contribute to the curvature perturbation
$\zeta$ either through the end-of-inflation mechanism \cite{soda} or directly,
if it acts as a vector curvaton after the end of inflation
\cite{vecurv,vecrev}. We have looked into the latter possibility but we have
found that the contribution to statistical anisotropy in $\zeta$ becomes
important only when the gauge coupling is fine-tuned to unnaturally small
values. This is because the attractor solution reached by the system during
inflation is such that the vector field contribution to the energy density is
rather small. As a result, a long period of vector field oscillations is
required after the end of inflation for the vector field to become a
significant fraction of the density budget. This can be achieved only at the
expense of a small gauge coupling. For realistic values of the gauge coupling
the decay rate of the vector field is not very small which does not allow for a
large period of vector field oscillations after inflation. However, this does
not mean necessarily that the vector curvaton mechanism is hopeless in
generating observable statistical anisotropy in this model. Indeed, the above
problem may be overcome if the magnitude of the vector field condensate is
increased at the end of inflation due to parametric resonance effects. We are
investigating this possibility in ref.~\cite{future} and so far the results are
promising.

The vector field can generate statistical anisotropy through other mechanisms
as well. For example, statistical anisotropy in $\zeta$ can be also due to the
mild anisotropisation of the Universe expansion \cite{anisinf,anisinf+}, also a
feature of the vector scaling solution \cite{Wagstaff}.
The end-of-inflation mechanism is a potentially more efficient way to introduce
statistical anisotropy in $\zeta$ due to the vector field perturbations. Since
the vector field is coupled to the GUT Higgs field, its perturbations modulate the
effective mass of this Higgs field. This, in turn, modulates the critical value of the
inflaton, which triggers the GUT phase transition and terminates inflation
\cite{soda}. However, the vector field contribution to the mass of the Higgs
field depends on the zero-mode value of the vector field during inflation, i.e.
{\em before} the phase transition. It is not clear how to evaluate this
because, during inflation, the vector field is massless (the GUT symmetry is
restored) so the theory is gauge invariant. This means that one can always
perform a gauge transformation of the form
\mbox{$\mbox{\boldmath $A$}\rightarrow\mbox{\boldmath $A$}+
\mbox{\boldmath $C$}$}, where \mbox{\boldmath $C$} is a constant spatial vector
of arbitrary magnitude. Thus, the value \mbox{\boldmath $A$} of the zero-mode
is gauge-dependent, i.e. it is not simply an environmental quantity as is the
expectation value of the scalar curvaton during inflation.
Finally, another method to introduce statistical anisotropy in $\zeta$ by the
vector field is through inhomogeneous reheating, since the effective mass of
the GUT Higgs (which is modulated by the vector field) also determines its
decay rate. For the varying kinetic function model, this method was touched
upon in ref.~\cite{sugravec}, where it was not found promising.

All in all we have shown that all the pathologies associated with the infamous
$\eta$-problem of inflation in supergravity are eliminated when considering an
interaction between the inflaton and a gauge field of the theory. Since in
supergravity the gauge kinetic function is a holomorphic function of the scalar
fields of the theory, it is quite natural to expect its modulation due to the
variation of the inflaton. Thus, the appearance of our so-called vector scaling
solution, is a rather generic phenomenon in supergravity theories with a gauge
field content. As we have shown, this scaling solution not only allows steep
inflation but also generates a weakly red spectrum of curvature perturbations
in agreement with observations.


\acknowledgments
This work is (in part) supported by the European Union under the Marie Curie Initial Training Network "UNILHC" PITN-GA-2009-237920. KD was supported (in part) by the Lancaster-Manchester-Sheffield Consortium for Fundamental Physics under STFC grant ST/J000418/1. JMW is supported by the Lancaster University Physics Department. KD wishes to thank the University of Crete for the hospitality.


\begin{thebiblio}{03}

\bibitem{book}
A.~R.~Liddle and D.~H.~Lyth,
{\it The Primordial Density Perturbation: Cosmology, Inflation and the origin of Structure}
(Cambridge Univ. Press, Cambridge U.K., 2009).

\bibitem{eternal}
A.~D.~Linde,
  Mod.\ Phys.\ Lett.\  A {\bf 1} (1986) 81.

\bibitem{hybrid}
A.~D.~Linde,
  Phys.\ Lett.\  B {\bf 259} (1991) 38;
  Phys.\ Rev.\  D {\bf 49} (1994) 748;
  Phys.\ Lett.\  B {\bf 249} (1990) 18;
F.~C.~Adams, K.~Freese,
  Phys.\ Rev.\  D {\bf 43} (1991) 353.

\bibitem{susyhybrid}
G.~R.~Dvali, Q.~Shafi and R.~K.~Schaefer,
  Phys.\ Rev.\ Lett.\  {\bf 73}, 1886 (1994);
G.~Lazarides, R.~K.~Schaefer and Q.~Shafi,
  Phys.\ Rev.\  D {\bf 56} (1997) 1324.

\bibitem{randall}
M.~Dine, L.~Randall and S.~Thomas,
Nucl.\ Phys.\ B {\bf 458} (1996) 291;
Phys.\ Rev.\ Lett.\  {\bf 75} (1995) 398;
D.~H.~Lyth and T.~Moroi,
  JHEP {\bf 0405} (2004) 004.

\bibitem{sugrahybrid}
E.~J.~Copeland, A.~R.~Liddle, D.~H.~Lyth, E.~D.~Stewart, D.~Wands,
  Phys.\ Rev.\  D {\bf 49} (1994) 6410.

\bibitem{double}
J.~Silk and M.~S.~Turner,
  Phys.\ Rev.\  D {\bf 35} (1987) 419.

\bibitem{extended}
D.~La, P.~J.~Steinhardt,
  Phys.\ Rev.\ Lett.\  {\bf 62} (1989) 376
  [Erratum-ibid.\  {\bf 62} (1989) 1066].

\bibitem{assisted}
A.~R.~Liddle, A.~Mazumdar, F.~E.~Schunck,
  Phys.\ Rev.\  D {\bf 58}, 061301 (1998);
P.~Kanti, K.~A.~Olive,
  Phys.\ Rev.\  D {\bf 60}, 043502 (1999);
E.~J.~Copeland, A.~Mazumdar, N.~J.~Nunes,
  Phys.\ Rev.\  D {\bf 60}, 083506 (1999).

\bibitem{Nflation}
S.~Dimopoulos, S.~Kachru, J.~McGreevy, J.~G.~Wacker,
  JCAP {\bf 0808}, 003 (2008).

\bibitem{curv}
D.~H.~Lyth and D.~Wands,
  Phys.\ Lett.\  B {\bf 524}, 5 (2002);
K.~Enqvist and M.~S.~Sloth,
Nucl.\ Phys.\ B {\bf 626} (2002) 395;
T.~Moroi and T.~Takahashi,
Phys.\ Lett.\ B {\bf 522} (2001) 215
[Erratum-ibid.\ B {\bf 539} (2002) 303].

\bibitem{liber}
 K.~Dimopoulos, D.~H.~Lyth,
  Phys.\ Rev.\  D {\bf 69}, 123509 (2004).

\bibitem{natural}
K.~Freese, J.~A.~Frieman and A.~V.~Olinto,
Phys.\ Rev.\ Lett.\  {\bf 65} (1990) 3233;
F.~C.~Adams, J.~R.~Bond, K.~Freese, J.~A.~Frieman and A.~V.~Olinto,
Phys.\ Rev.\  D {\bf 47} (1993) 426;
L.~Knox and A.~Olinto,
Phys.\ Rev.\  D {\bf 48} (1993) 946;
K.~Freese and W.~H.~Kinney,
Phys.\ Rev.\  D {\bf 70} (2004) 083512.

\bibitem{lorenzo}
M.~M.~Anber and L.~Sorbo,
  Phys.\ Rev.\  D {\bf 81} (2010) 043534.

\bibitem{pngb}
K.~Dimopoulos, D.~H.~Lyth, A.~Notari, A.~Riotto,
  JHEP {\bf 0307}, 053 (2003).

\bibitem{Wagstaff}
  J.~M.~Wagstaff, K.~Dimopoulos,
  Phys.\ Rev.\  D {\bf 83}, 023523 (2011).

\bibitem{vecurv}
K.~Dimopoulos,
Phys.\ Rev.\  D {\bf 74} (2006) 083502.

\bibitem{stanis}
K.~Dimopoulos, M.~Kar\v{c}iauskas, D.~H.~Lyth and Y.~Rodriguez,
  JCAP {\bf 0905} (2009) 013.

\bibitem{soda}
  S.~Yokoyama and J.~Soda,
  JCAP {\bf 0808} (2008) 005.

\bibitem{AoE}
K.~Land and J.~Magueijo,
  Phys.\ Rev.\ Lett.\  {\bf 95} (2005) 071301;
Mon.\ Not.\ Roy.\ Astron.\ Soc.\ (2007) {\bf 378} 153.

\bibitem{GE}
 N.~E.~Groeneboom and H.~K.~Eriksen,
  Astrophys.\ J.\  {\bf 690} (2009) 1807;
N.~E.~Groeneboom, L.~Ackerman, I.~K.~Wehus and H.~K.~Eriksen,
  Astrophys.\ J.\  {\bf 722}, 452 (2010);
D.~Hanson and A.~Lewis,
  Phys.\ Rev.\  D {\bf 80} (2009) 063004;
Y.~-Z.~Ma, G.~Efstathiou and A.~Challinor,
  Phys.\ Rev.\  D {\bf 83}, 083005 (2011).

\bibitem{bartolo}
E.~Dimastrogiovanni, N.~Bartolo, S.~Matarrese and A.~Riotto,
  Adv.\ Astron.\  {\bf 2010}, 752670 (2010).

\bibitem{fnlanis}
 M.~Kar\v{c}iauskas, K.~Dimopoulos and D.~H.~Lyth,
  Phys.\ Rev.\  D {\bf 80} (2009) 023509.

\bibitem{mota}
S.~Hervik, D.~F.~Mota and M.~Thorsrud,
  arXiv:1109.3456.

\bibitem{anisinf}
S.~Kanno, M.~Kimura, J.~Soda and S.~Yokoyama,
  JCAP {\bf 0808} (2008) 034;
  M.~-a.~Watanabe, S.~Kanno and J.~Soda,
  Phys.\ Rev.\ Lett.\  {\bf 102} (2009) 191302.

\bibitem{anisinf+}
C.~Pitrou, T.~S.~Pereira and J.~-P.~Uzan,
  JCAP {\bf 0804}, 004 (2008);
M.~-a.~Watanabe, S.~Kanno and J.~Soda,
  Mon.\ Not.\ Roy.\ Astron.\ Soc.\  {\bf 412}, L83 (2011);
M.~-a.~Watanabe, S.~Kanno and J.~Soda,
  Prog.\ Theor.\ Phys.\  {\bf 123} (2010) 1041;
T.~R.~Dulaney and M.~I.~Gresham,
  Phys.\ Rev.\  D {\bf 81} (2010) 103532;
A.~E.~Gumrukcuoglu, B.~Himmetoglu and M.~Peloso,
  Phys.\ Rev.\  D {\bf 81} (2010) 063528;
B.~Himmetoglu,
  JCAP {\bf 1003} (2010) 023.

\bibitem{varkin}
K.~Dimopoulos, M.~Kar\v{c}iauskas and J.~M.~Wagstaff,
  Phys.\ Rev.\  D {\bf 81} (2010) 023522;
  Phys.\ Lett.\  B {\bf 683} (2010) 298.

\bibitem{sugravec}
K.~Dimopoulos,
  Phys.\ Rev.\  D {\bf 76} (2007) 063506.

\bibitem{nonmin}
K.~Dimopoulos and M.~Kar\v{c}iauskas,
  JHEP {\bf 0807} (2008) 119.

\bibitem{wald}
  R.~M.~Wald,
  Phys.\ Rev.\  D {\bf 28}, 2118 (1983).

\bibitem{sodanew}
S.~Kanno, J.~Soda and M.~-a.~Watanabe,
  JCAP {\bf 1012}, 024 (2010).


\bibitem{vecrev}
K.~Dimopoulos,
  arXiv:1107.2779.

\bibitem{wmap}
E.~Komatsu {\it et al.} [ WMAP Collaboration ],
  Astrophys.\ J.\ Suppl.\  {\bf 192}, 18 (2011).

\bibitem{planck}
 A.~R.~Pullen and M.~Kamionkowski
 Phys.\ Rev.\  D {\bf 76} (2007) 103529.

\bibitem{dN}
A.~A.~Starobinsky,
  JETP Lett.\  {\bf 42}, 152 (1985)
  [Pisma Zh.\ Eksp.\ Teor.\ Fiz.\  {\bf 42}, 124 (1985)];
M.~Sasaki and E.~D.~Stewart,
  Prog.\ Theor.\ Phys.\  {\bf 95}, 71 (1996);
D.~H.~Lyth, K.~A.~Malik and M.~Sasaki,
  JCAP {\bf 0505}, 004 (2005).

\bibitem{cesar}
C.~A.~Valenzuela-Toledo, Y.~Rodriguez and D.~H.~Lyth,
  Phys.\ Rev.\  D {\bf 80} (2009) 103519;
C.~A.~Valenzuela-Toledo and Y.~Rodriguez,
  Phys.\ Lett.\  B {\bf 685} (2010) 120.

\bibitem{Lazarides:2008nx}
  G.~Lazarides, I.~N.~R.~Peddie and A.~Vamvasakis,
  Phys.\ Rev.\  D {\bf 78}, 043518 (2008).

\bibitem{Lazarides:1985bj}
G.~Lazarides, C.~Panagiotakopoulos and Q.~Shafi,
  Phys.\ Rev.\ Lett.\  {\bf 56}, 432 (1986).

\bibitem{classifications}
  N.~Ganoulis, G.~Lazarides and Q.~Shafi,
  Nucl.\ Phys.\  B {\bf 323} (1989) 374;
  G.~Lazarides and Q.~Shafi,
  Nucl.\ Phys.\  B {\bf 329} (1990) 182.

\bibitem{Lazarides:1998zf}
  G.~Lazarides and N.~Tetradis,
  Phys.\ Rev.\  D {\bf 58} (1998) 123502.

\bibitem{Coleman:1973jx}
S.~R.~Coleman and E.~J.~Weinberg,
  Phys.\ Rev.\  D {\bf 7}, 1888 (1973).

\bibitem{origin}
G.~Lazarides and Q.~Shafi, Phys.\ Lett.\ B {\bf 258}, 305 (1991).

\bibitem{direct} 	
T.~Dent, G.~Lazarides, and R.~Ruiz de Austri,
Phys.\ Rev.\ D {\bf 69}, 075012 (2004);
Phys.\ Rev.\ D {\bf 72}, 043502 (2005).

\bibitem{future}
K.~Dimopoulos, G.~Lazarides and J.~M.~Wagstaff, in preparation.



\end{thebiblio}
\end{document}